\documentclass{aa}  

\usepackage{graphicx}
\usepackage{tikz}

\usepackage{txfonts}
\usepackage{ulem}   
\usepackage[colorlinks=true,linkcolor=blue,citecolor=blue]{hyperref}
\usepackage{multirow}
\usepackage{float}

\definecolor{mpc}{rgb}{0.2, 0.20, 0.80}
\definecolor{GR}{rgb}{0.8, 0.20, 0.20} 
\definecolor{GRc}{rgb}{0.20, 0.20, 0.50}

\newcommand{\Herschel}{\textit{Herschel} }
\newcommand{\Spitzer}{\textit{Spitzer} }
\newcommand{\EN}{ELAIS~N1\,}
\newcommand{\mstar}{$M_{star}$\,}
\newcommand{\mdust}{$M_{dust}$\, }
\newcommand{\ldust}{$L_{dust}$\,}

\newcommand{\AFUV}{$A_{FUV}$\,}

\defcitealias{Bethermin2017}{B17}
\defcitealias{Jana2020}{BB20}

\begin{document} 

   \title{Getting ready for the LSST data - estimating the physical properties of $z<2.5$ main sequence galaxies }

   \subtitle{}

 \author{G. Riccio\inst{1}  \and K. Małek\inst{1,2} \and A. Nanni\inst{1,2} \and M. Boquien\inst{4} \and V. Buat\inst{2} \and D. Burgarella\inst{2} \and D. Donevski\inst{3} \and M. Hamed\inst{1}  \and P. Hurley\inst{6}  \and R. Shirley\inst{6,7} \and A. Pollo\inst{1,5}}

    \institute{National Centre for Nuclear Research, ul. Pasteura 7, 02-093 Warszawa, Poland
    \and
    Aix Marseille Univ. CNRS, CNES, LAM, Marseille, France
    \and
    SISSA, Via Bonomea 265, Trieste, Italy
    \and
    Centro de Astronomía (CITEVA), Universidad de Antofagasta, Avenida Angamos 601, Antofagasta, Chile
    \and
    Astronomical Observatory of the Jagiellonian University, ul. Orla 171, 30-244 Cracow, Poland
    \and
    Astronomy Centre, Department of Physics and Astronomy, University of Sussex, Falmer, Brighton BN1 9QH, UK
    \and
    Instituto de Astrofísica de Canarias, E-38205 La Laguna, Tenerife, Spain}

 
  \abstract
   {}
   { In this work we study how to employ the upcoming \textit{Legacy Survey of Space and Time} (LSST) data, from the Vera C.~Rubin Observatory, to constrain physical properties of normal, star forming galaxies (so-called, the main sequence galaxies). 
   Since the majority of observed LSST objects will have no auxiliary data, we use  simulated LSST data and existing real observations to test the reliability of estimations of the physical properties of galaxies, such as their star formation rate (SFR), stellar mass (\mstar), and dust luminosity (\ldust).
   We focus on normal star-forming galaxies, as they form the majority of the galaxy population in the universe and therefore are more likely to be observed by the LSST. }
   {We perform a simulation of LSST observations and uncertainties of 50\,385 real galaxies within redshift range $0<z<2.5$. 
   In order to achieve this goal, we used the unique multi-wavelength data from the  \textit{Herschel Extragalactic Legacy Project} (HELP) survey. 
   Our analysis focus on two fields:  \EN and COSMOS. 
   To obtain galaxy physical properties we fit their Spectral Energy Distributions (SEDs) using the Code Investigating GALaxy Emission (CIGALE). 
   We simulate the LSST data by convolving the SEDs fitted by employing the multi-wavelength observations.
   We compare the main galaxy physical properties, such a SFR, \mstar, and \ldust  obtained from the fit of the observed multi-wavelength photometry of galaxies (from ultraviolet  to far~infrared) to the ones obtained from the simulated LSST optical measurements only. }
   {We present the catalogue of simulated LSST observations for 23\,291 main sequence  galaxies in the \EN field and 9\,093 in the COSMOS field, available on the HELP virtual observatory. 
   The stellar masses estimated based on the LSST measurements are in agreement with the  full UV-FIR  SED estimations, as they depend mainly on the UV and optical emission, well covered by LSST in the considered redshift range.
   Instead we obtain a clear overestimation of the dust related properties (SFR, \ldust, \mdust) estimated with LSST only, highly correlated with redshift. 
   We investigate the cause of this overestimation and we conclude that it  is related to an overestimation of the dust attenuation, both UV and NIR. 
   We find that it is necessary to employ auxiliary rest-frame mid-infrared observations, simulated UV observations, or FUV attenuation (AFUV)-\mstar relation, to correct the overestimation. 
   We also deliver the correction  formula  $\rm \log_{10}({SFR_{LSST}}/{SFR_{real}})=0.26\cdot \textit{z}^2 -0.94\cdot \textit{z} +0.87$, based on the 32\,384 MS galaxies with \Herschel detection.
   }
   {}

    
   \maketitle

\section{Introduction}
\label{Sec:Introduction}

 In the last 20 years, the study of the multi-wavelength emission of galaxies from X-rays to radio was found to be necessary to properly analyse the physical properties of galaxies. As the spectral energy distribution (SED) is the result of a complex interplay between several components: old and young stars, stellar remnants, interstellar medium, dust and supermassive black holes (\citealt{Walcher2011}, \citealt{Conroy2013}), only the  panchromatic view of galaxies can give the full information about their physical properties. For example, the emission from the hot interstellar medium, active galactic nuclei (AGN) or stellar remnant can be observed in the X-ray band \citep{Fabbiano2006}, while the emission of the dust heated by the interstellar radiation can be observed in mid and far-infrared band (\citealt{Silva1998}, \citealt{Noll2009}, \citealt{daCunha2010}, \citealt{Hao2011}, \citealt{Calzetti2012}, \citealt{Schreiber2018}, \citealt{Leja2018}). To fully comprehend the interactions between these parts, the simultaneous use of different spectral ranges is needed. As broad band photometry is much less expensive than spectroscopy, in terms of observation time, modelling the broad band SED of galaxies has become one of the most common methods to evaluate and constrain the  physical properties. In this way, properties such as star formation rate (SFR) and stellar mass \mstar, that are essential to have a complete understanding of galaxy formation and evolution, can be evaluated.

However, modelling the SED can be an intricate problem, as galaxies with very different properties can look similar over some wavelength range: i.e. a young dusty galaxy can imitate old dust-free galaxy, as both look red in the optical. This is particularly the case when considering restricted wavelength ranges rather than the full SED, that is rarely available. Therefore, estimating the physical properties with only a limited wavelength range is an important challenge for SED modelling.

In the literature (i.e. \citealt{Kennicutt98},  \citealt{LeFloc2005}, \citealt{Schreiber2015}, \citealt{Whitaker2017}) it has already been shown that the ultraviolet (UV) to infrared (IR) SED contains important information about the star formation activity of galaxies. For example, some knowledge on newborn stars can be directly inferred from the UV band, making it a very efficient tracer of the SFR. But, the region were these stars are created are highly obscured by dust, making them very difficult to observe. Dust, composed of carbonaceous and silicate grains, absorbs part of the UV emission and re-emits it in the IR band. 
For example, Fig. 10 of \cite{Buat2019} shows that the total SFR of a galaxy is a sum of the SFR obtained from UV/optical measurements and the SFR estimated from IR data. Therefore, considering the important role played by dust, the introduction of attenuation laws, that describe how dust obscures the light coming from the stars, is fundamental for the SED fitting process. 

The attenuation law built by \cite{Calzetti1994} for nearby UV-bright starburst galaxies is by far the most commonly used in literature. However other laws such as the one proposed by \cite{charlot2000} and \cite{LoFaro2017} are widely employed in the SED fitting codes. \cite{Malek2018} used a combination of UV and IR observations to find the best approach to fit SEDs of million of galaxies from the \textit{Herschel Extragalactic Legacy Project} (HELP) across a wide redshift range ($0<z<6$) to obtain homogeneous estimates of the main physical properties. They found that, using three different attenuation laws, the estimation of stellar masses can change by a factor of 2 on average. Similar results were found, e.g., in a sample of Ultra Luminous IR Galaxies at $z\sim2$ by \cite{LoFaro2017}, for galaxies obtained from the semi-analytic galaxy formation model GALFORM  by \cite{Mitchell2013}, and by \cite{Burgarella2013} who combined UV to IR measurement up to $z=3.6$ to calculate the redshift evolution of the total SFR and dust attenuation. They found that the attenuation increases up to $z=1.2$ and then decreases at higher redshift. Also the ratio between UV and far~IR (FIR) emission serves as an indicator of the dust attenuation in galaxies (\citealt{Buat2005}, \citealt{Takeuchi2005}). All these factors make the combined usage of UV and IR observations necessary to provide a better understanding of the star formation history (SFH), SFR and dust attenuation properties of the galaxies. In order to perform the SED fitting of galaxies, different methods and codes were developed, such as STARLIGHT \citep{CidFernandes2005}, VESPA \citep{Tojeiro2007}, Hyperz \citep{Bolzonella2000}, Le Phare (\citealt{Arnouts1999}, \citealt{Ilbert2006}),  PÉGASE.3 (\citealt{Fioc2019}), COSMOS2020 (\citealt{Weaver2021}) together with bayesian SED fitting codes such as GOSSIP \citep{Franzetti2008}, PROSPECTOR \citep{Leja}, CIGALE (\citealt{Noll2009}; \citealt{Boquien2019}), BayeSED \citep{Han2014} .

The main problem of the multi-wavelength fitting technique is the lack of high quality IR observations, both due to instrumental sensitivity and lower resolution of wide IR cameras or extremely expensive sub-millimeter observations, i.e.  ALMA. On the contrary, a very large and high quality coverage of the optical part of the spectrum is usually available, both for wide field surveys and for narrow and deep field imaging, thanks to several ground and space telescopes. With the upcoming \textit{Legacy Survey of Space and Time} (LSST, \citealt{Ivezic2019}) from the Vera C.~Rubin Observatory,  we will obtain even higher quality optical images in the \textit{ugrizy} bands. The LSST survey will observe around 20~billion galaxies during 10 years of observations. Most of this galaxies will not have any counterpart in the available IR catalogues. Moreover, IR astronomy often suffers from blending issues, which makes  precise matching between optical and IR sources even more difficult (\citealt{Hurley2017}, \citealt{Pearson2018}). 

The LSST will be the largest (8.4 meters of primary mirror) wide field ground telescope designed to obtain repeated images covering the sky visible from Cerro Pachón in Chile. The survey will observe around 30\,000 $deg^2$ of the southern sky, covering the wavelength range 320-1\,050 nm. It will reach, at the end of the 10 years survey, a magnitude depth $\sim 27.5$ in $r$ band and similar in the other bands. 
Considering the depth of forthcoming observations, it is expected that LSST will unveil a significant number of faint galaxies that have remained undetected in current wide area surveys. These potentially large datasets will raise multifold questions such as: how can we use only LSST optical observations to obtain estimates of the main physical properties of galaxies? How realistic and reliable would they be?
In this paper we investigate the aforementioned topics by performing a simulation of LSST observations of main-sequence (hereafter MS) galaxies which form a nearly linear relation (in log-log space) between their stellar mass and SFR (\citealt{Noeske2007}, \citealt{Elbaz2010}, \citealt{Rodighiero2011}, \citealt{Speagle2014}, \citealt{Schreiber2015}, \citealt{Whitaker2015}, \citealt{Pearson2018}).
Main sequence galaxies constitute the dominant population in deep fields such as COSMOS and \EN that can reach very faint optical magnitudes ($\sim$ 29-30 mag). LSST is expected to expand, among others, the observed MS population of galaxies to other fields which are not currently covered by deep field surveys. For this reason, we decided to focus on the MS galaxies.

The paper is organised as follows. In Section~\ref{Sec:Data} we describe the data and the HELP project. In Section~\ref{Sec:Methodology}  we present the sample selection, outliers and starburst, and the methodology used for this work. In Section~\ref{Sec:Results}  we discuss the simulated LSST magnitude and errors. The same Section, together with Section \ref{different_param} and \ref{Sec:AFUVMSTAR}, presents the results. Our conclusions are presented in Section~\ref{Sec:Conclusions}. 
Throughout this paper we use WMAP7 cosmology \citep{Komatsu2011}: $\Omega_m = 0.272$,
$\Omega_{\Lambda}$ = 0.728, $H_0$ = 70.4 km s$^{-1}$ Mpc$^{-1}$.

\section{Data}
\label{Sec:Data}

The HELP collaboration provides extremely valuable multi-wavelength data over the HerMES \citep{Oliver2012} and the H-ATLAS survey fields \citep{Eales2010} and other relevant \Herschel fields. The total area of HELP is 1\,269.1 deg.$^2$ (Oliver et al. in prep, \citealt{Shirley2019}). 
\textit{Herschel} was equipped with two imaging instruments, the Photodetector Array Camera and Spectrometer (PACS; \citealt{Poglitsch2010}), that observed the FIR  at 100 and 160 $\mu m$, and the Spectral and Photometric Imaging Receiver (SPIRE; \citealt{Griffin2010}) that covered 250, 350, and 500~$\mu m$ wavelength ranges.

Surveys that combine a wide range of wavelengths have particular identification issues due to the different spatial resolution of the sources in different bands. To correct this issue HELP builds a master list catalogue of objects as complete as possible for each field and uses the NIR sources of this catalogue as prior information to deblend the \Herschel maps. Detailed description can be found in \cite{Shirley2019}. The tool developed to obtain the photometry of \Herschel sources,  XID+ \citep{Hurley2017}, is a probabilistic de-blending algorithm which extracts source flux densities from photometry maps that suffer from source confusion. It uses Bayesian inference to explore the posterior probability distribution and provide probability density function (PDFs) for all prior sources, and thus flux and uncertainties can be estimates. A detailed description can be found in \cite{Hurley2017}.
The whole procedure is described in Notebook and stored in a GitHub repository\footnote{\url{https://github.com/H-E-L-P/dmu_products}}.

In the following paper, we use two HELP fields, due to the wealth of multi-wavelength data available within their wide field coverage: the European Large Area ISO Survey North~1, hereafter \EN \citep{Oliver2000} and COSMOS field \citep{Laigle2016}. 
Besides data from two PACS and three SPIRE maps, we used available sets of photometric data at shorter wevelenghts for both fields (listed in Table~\ref{table:1} and described below in Sections~\ref{subsec:ELAIS}~and~\ref{subsec:COSMOS}).   
Based on true galaxy observations from that fields we evaluate the 'LSST-like' observations used for further analysis.

\subsection{\EN  field}
\label{subsec:ELAIS}
 According to the HELP strategy,  all  sources  detected in any of the Spitzer IRAC bands were used as a prior for XID+  to obtain FIR fluxes. 
 XID+ was run on the \Spitzer MIPS~24~$\mu$m and \Herschel PACS and SPIRE maps. 
 The flux level at which the average posterior probability distribution of the source flux becomes Gaussian is 20 $mJy$ for MIPS, 12.5 and 17.5~$mJy$ for 100~$\mu m$ and 160~$\mu m$ PACS bands respectively, and 4~$ mJy$ for all three (250, 350 and 500$\mu m$) SPIRE bands (see for more details \citealt{Hurley2017}; indicating that the information from data dominates over the prior). In the de-blending procedure, the priors used for computing the fluxes satisfied two criteria: they must have an IRAC~1 band detection and they must have been detected in either the optical or near IR (NIR) wavelengths to eliminate artefacts. 
 More information about the catalogue can be found in \cite{Malek2018}, Shirley et al. (in prep.) and on the main webpage of the HELP project\footnote{\url{http://hedam.lam.fr/HELP/}}.

In addition to FIR bands, the catalogue is built on a position cross match of all the public survey data available in the optical and mid~IR (MIR) range. This comprises observation from \textit{Isaac Newton} Telescope/Wide Field Camera (INT/WFC) survey \citep{Gonzalez2011}, the Subaru Telescope/Hyper Suprime-Cam Strategic Program Catalogues (HSC-SSP; \citealt{Aihara2018}), the Panoramic Survey Telescope and Rapid Response System (Pan-STARRS; \citealt{Chambers2016}), the UK Infrared Telescope Deep Sky Survey – Deep Extragalactic Survey (UKIDSS–DXS) (\citealt{Swinbank2013}; \citealt{Lawrence2007}), the \textit{Spitzer} Extragalactic Representative Volume Survey (SERVS; \citealt{Mauduit2012}), and the \textit{Spitzer} Wide InfraRed Extragalatic survey (SWIRE; \citealt{Lonsdale2003}; \citealt{Stauffer2005}). We show the list of filters for \EN in Table~\ref{table:1}. 
The whole matching procedure is described in \cite{Shirley2019}.

\subsection{COSMOS field}
\label{subsec:COSMOS}

For the COSMOS field, the XID+ 
analysis was perfomed on \Spitzer and \Herschel maps for all the sources with fluxes greater than 1~$\mu Jy$ in any of the IRAC bands from the COSMOS2015 catalogue \citep{Laigle2016}. 
The fluxes obtained follow the criterion of goodness defined in XID+ and corresponds to a Gaussian posterior distribution of the estimated flux.

Starting with this multi-wavelength catalogue, COSMOS2015 \citep{Laigle2016}, ancillary photometry is added with a position cross match with other public survey containing, again, optical and MIR observations. 
This comprises, other than the one already mentioned for \EN, the WIRCam Deep Survey (WIRDS, WIRcam bands J, H, Ks), the VLT Survey Telescope (VST; \citealt{Arnaboldi1998}), the Victor Blanco 4-m Telescope, the Visible and Infrared Survey Telescope for Astronomy (VISTA; \citealt{Emerson2006}, \citealt{Dalton2006}) and UKIDSS-LAS (WFCAM bands J, H, K) catalogues. The merging strategy is the same as for \EN and is described in detail in \citealt{Shirley2019}). The list of filters used for COSMOS survey is shown in Table~\ref{table:1}.

Detailed description of both fields (the area, mean depths in different filters, all raw files and ancillary data and many others)  can be found on the  \url{http://hedam.lam.fr/HELP/dataproducts/dmu31/dmu31_Field_overviews/} webpage.

\begin{table*}
\caption{List of filter for \EN and COSMOS fields. }   
\label{table:1}      
\centering                          
\renewcommand{\arraystretch}{1.1} 
\begin{tabular}{c c c c}        
\hline\hline                 
Telescope & Instrument & Elais-N1 filters & COSMOS filters \\    
\hline                        
\multirow{2}{*}{CFHT} & MegaCam & u, g, r, y, z & u, g, r, i, y, z  \\
& WIRcam & & H, J, Ks\\ \hline
Subaru & HSC & g, r, i, z, N921, y & g, r, i, z, N921, y \\ \hline
Isaac Newton & Wide Field Cam. & u, g, r, i, z \\
PanSTARRS1 & Gigapixel Cam.1 & g, r, i, z, y & g, r, i \\\hline
UKIRT & WFCam & J, K & J, H, K \\ \hline
VST & OmegaCAM & & u,g,r,i \\ \hline
BLANCO & DEcam & & g, r, z\\ \hline
VISTA & Vircam & & J, H, Ks, y\\ \hline
\multirow{2}{*}{\Spitzer}  & IRAC & 3.6, 4.5, 5.8, 8.0 ( $\mu m$) &  3.6, 4.5,5.8, 8.0 ($\mu m$) \\
 & MIPS &  24 ($\mu m$) & 24 ($\mu m$) \\ \hline
\multirow{2}{*}{\Herschel} & PACS & 100, 160 ($\mu m$) & 100, 160 ($\mu m$) \\
 & SPIRE & 250, 350, 500 ($\mu m$) & 250, 350, 500 ($\mu m$) \\
 \hline
\end{tabular}
\vspace{1mm}

\end{table*}

\subsection{Total sample} 
\label{subsec:Final sample}

As part of the HELP database, both fields catalogues include photometric redshifts generated using a template fitting method, based on a Bayesian combination approach, described in \cite{Duncan2018}. They investigated the performance of three photometric redshift template sets as a function of redshift, radio luminosity and infrared/X-ray properties, over the NOAO Deep Wide Field Survey Bootes and COSMOS fields. The three template sets used are: (1) default EZY reduced galaxy set \citep{Brammer2008}, (2) “XMM COSMOS” templates \citep{Salvato2009}, and (3) atlas of Galaxy SEDs \citep{Brown2014}.

The total sample includes 39\,329 objects for the ELAIS N1 survey and 14~864 for COSMOS, with FIR detections in at least 
two photometric bands with signal-to-noise ratio ($S/N$)$\geqslant$3. This cut is performed to remove objects with unreliable photometry and thus improve the quality of the SED fitting process. We keep in mind that by employing the aforementioned selection we are restricting our analysis to only a subsample of galaxies that LSST will observe, that are objects bright in the FIR. Then, as we will show in the next Section, we have selected only the so-called MS galaxies observed in the spectral range from UV to FIR, as these are the most common type of galaxies observed. The considered bands are u, g, r, i, z, N921, y, J, H, K, Spitzer IRAC 3.6, 4.5, 5.8 and 8.0~$\mu m$, Spitzer MIPS 24~$\mu m$ and five passbands from Herschel, two from  PACS (100 and 160 $\mu m$) and three from SPIRE (250, 350 and 500~$\mu m$), across the ELAIS~N1 and COSMOS fields. 

\section{Methodology: SED fitting, starbursts and outliers detection}
\label{Sec:Methodology}
\subsection{SED fitting with CIGALE}
\label{Subsec:CIGALE}

The SED fitting is performed with the Code Investigating GALaxy Emission\footnote{\url{https://cigale.lam.fr}} (CIGALE) tool. 
For a detailed description of the code we refer to \cite{Boquien2019}, here we will give just a brief summary. 
CIGALE is a Bayesian SED fitting code that combines modelled stellar spectra with dust attenuation and emission. 
CIGALE preserves the energy balance considering both the energy emitted by massive stars, partially absorbed by dust grains and then re-emitted in the MIR and FIR.
The quality of the fit is expressed by the best $\chi^2$ (and a reduced best $\chi^2$ defined as $\chi^2_r=\chi^2/(N-1)$, with N the number of data points).  
The minimum value of $\chi^2_r$ corresponds to the best model selected from the grid of all possible computed models from the input parameters.
The physical properties and their uncertainties are estimated as the likelihood-weighted means and standard deviations. 

To obtain the starting MS sample of galaxies with the correct physical properties necessary to compare with those obtained with LSST only, we run CIGALE on the \EN and COSMOS samples with the physical modules and parameters reported in Table~\ref{table:appendix parameters}. 
We did not use the AGN module  (see App.~\ref{app:AGN} and Sec.~\ref{subsec:outlier}).
As shown in \cite{Malek2018}, this set of parameters corresponds to the one that best fits a large sample of IR detected galaxies in the 23 HELP fields, for the redshift range 0<z<6. 
We use a SFH modelled as delayed exponential function with an additional exponential burst to select and remove starburst galaxies from our sample in order to only retain MS galaxies. 
We perform the SED fitting by employing the modified version of  \cite{charlot2000} attenuation law, as already employed in \cite{Malek2018} for a large sample of multi-wavelength HELP data. 
We use the \cite{Draine2014} dust emission module. A detailed description of each module can be found in \cite{Boquien2019} and \cite{Malek2018}. 

\begin{table*}
\caption{Input parameter of the code CIGALE.}   
\label{table:appendix parameters}      
\centering                          
\renewcommand{\arraystretch}{1.1} 
\begin{tabular}{l| l}        
\hline                
Parameters & Values \\    

\hline
\multicolumn{2}{c}{Star formation history:} \\\hline\hline

       \textit{ Delayed star formation history + additional burst}\\ 
\hline
   e-folding time of the main stellar population model (Myr) & 1000, 2000, 3000, 5000, 7000 \\
   e-folding time of the late starburst population model (Myr) & 5000\\
   Mass fraction of the late burst population & 0.001, 0.01, 0.03, 0.1, 0.3 \\
   Age (Myr) & 1000, 2000, 3000, 4000, 5000, 6500, 10000\\
   Age of the late burst (Myr) & 10, 40, 70\\

\hline                        
        \textit{Delayed star formation history}\\ 
\hline
e-folding time of the main stellar population model (Myr) & 1000, 2000, 3000, 4000, 5000, 6500, 8000 \\
   Age (Myr) & 500, 1000, 3000, 4000, 5000, 6000, 7000, 8000, 9000,\\ & 10000, 12000\\
   Mass fraction of the late burst population & 0.0 \\
\hline                                   
\hline
\multicolumn{2}{c}{ Single stellar population  \cite{Bruzual2003}} \\\hline\hline

    Initial mass function & \cite{Chabrier2003}\\
    Metallicities (solar metallicity) & $0.02$\\
    Age of the separation between the young and the old star population (Myr) & $10$\\
\hline\hline
\multicolumn{2}{c}{ Dust attenuation law  \cite{charlot2000}} \\\hline\hline

    $A_V$ in the Birth Clouds & 0, 0.05, 0.1,  0.3, 0.8, 1.2, 1.7, 2.3, 2.8, 3.3, 3.8, 4.0, 4.2\\
    Power law slopes of the attenuation in the birth clouds & $-0.7$\\
    BC to ISM factor (Av ISM / Av BC) & 0.5, 0.8\\
    slope ISM & $-0.7$\\
\hline\hline
\multicolumn{2}{c}{Dust emission:}\\ \hline\hline

\textit{\cite{Draine2014} }\\
\hline

    Mass fraction of PAH & 1.12, 2.5, 3.19\\
    Minimum radiation field ($U_{min}$) & 5.0, 10.0, 25.0\\
    Power law slope dU/dM ($U^\alpha)$ & 2.0, 2.8\\
    Fraction illuminated from $U_{min}$ to $U_{max}$ ($\gamma$) & 0.02\\
\hline

\textit{\cite{Dale2014} }\\
\hline

    AGN fraction & 0 \\
    Power law slope dU/dM ($U^\alpha)$ & 2.0\\
\hline

\end{tabular}
\vspace{1mm}

\end{table*}

To improve the quality of our selection, from the full sample we only select objects with redshift lower than 2.5. 
Our cut in redshift is not related with the LSST redshift range, as the photometric redshifts for LSST will be applied and calibrated over the range 0 $<$ z $<$ 4 for galaxies to r$\sim$27.5 \citep{LSST_SB_2009}. 
The z$<$2.5 is related to the redshift distribution in the \EN and COSMOS fields used in this analysis, and also it restricts us to high quality data and so maximize the accuracy of the estimation of physical properties.
Moreover, we removed all the objects recognized as possible stars by GAIA  (\textit{flag GAIA$>$0} in the database). 
In this way we remove 2\,921 objects (7.5\% of the sample) from \EN and 887 (6\% of the sample) from the COSMOS catalogue. 

From now on, we will refer to the remaining 36\,408 and 13\,977 galaxies from the \EN and COSMOS fields respectively, as the real sample.

\subsection{Starburst galaxies selection}
\label{subsec:Starburst}
Galaxies can be classified according to their different properties: morphology, colour, environment, mass etc. 
One of the properties often used in the literature is the  rate of which stars are forming out of gas, the SFR.
That leads to define three different types of galaxies: passive, normal/MS, and starburst (SB). 
The boundaries dividing these classifications are not precisely defined, as different authors use different methods to distinguish starbursts from MS galaxies (i.e. \citealt{Rodighiero2011}, \citealt{Speagle2014}, \citealt{Elbaz2018}, \citealt{Darko2020}). 
A universally accepted method does not exist. 
Nevertheless, there is agreement that the three groups differ in regard to their evolution and physical properties, like SFH, dust and gas content, and others ({\citealt{Silverman2018}, \citealt{Elbaz2018}}). 

Most galaxies observed with LSST will be composed of active IR galaxies, but the majority of them are likely to be normal, MS--like, or passive galaxies. 
For example, the current estimate on SB galaxies contributing to the full star-formation population is about ~5\% (\citealt{Schreiber2016}, \citealt{Bethermin2017}). However, this contribution increases if we isolate brighter IR galaxies only (e.g. \citealt{Miettinen2017}).

\begin{figure*}[htb]
    \centering
    \includegraphics[width=0.5\hsize]{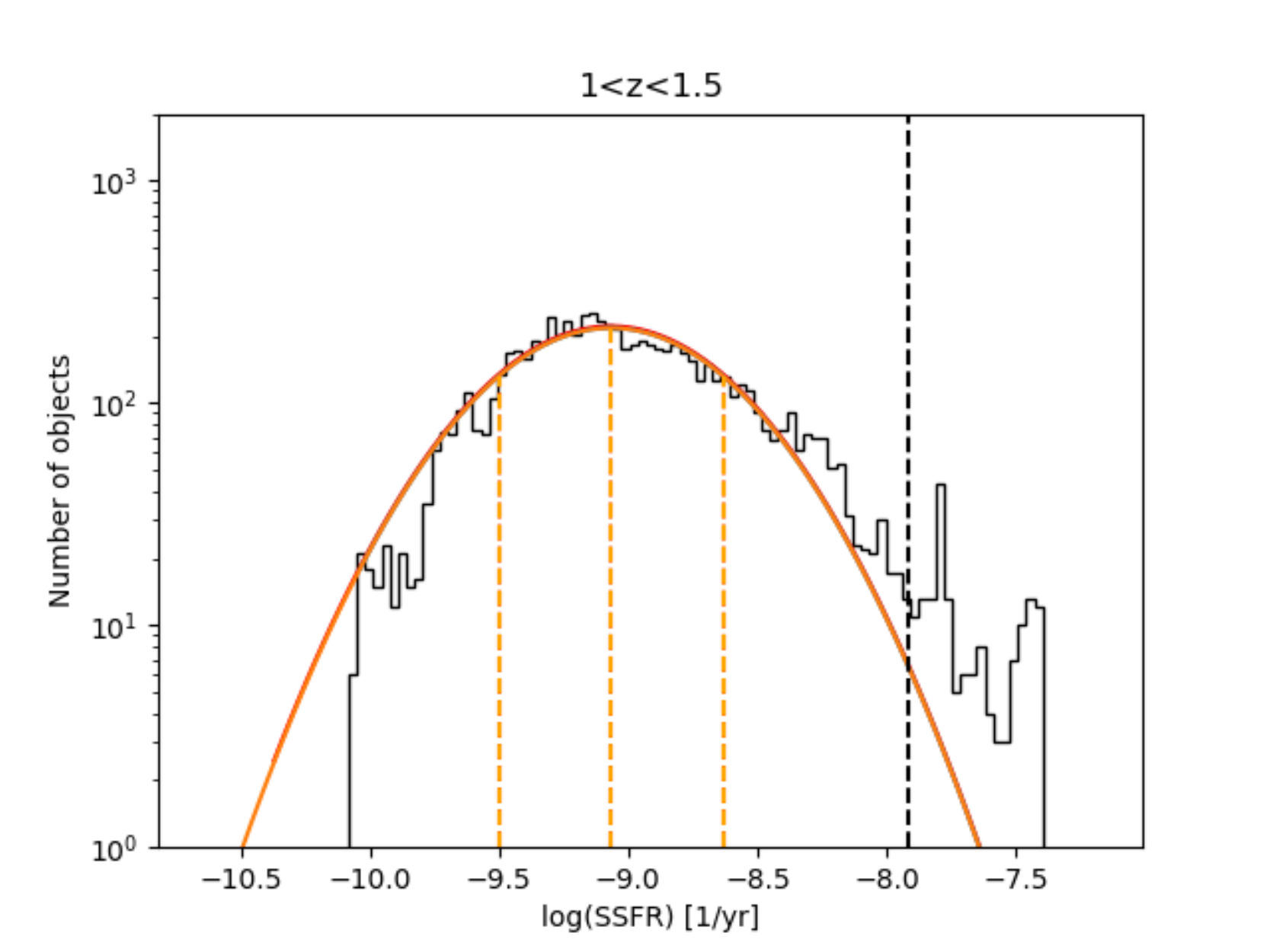}%
    \includegraphics[width=0.5\hsize]{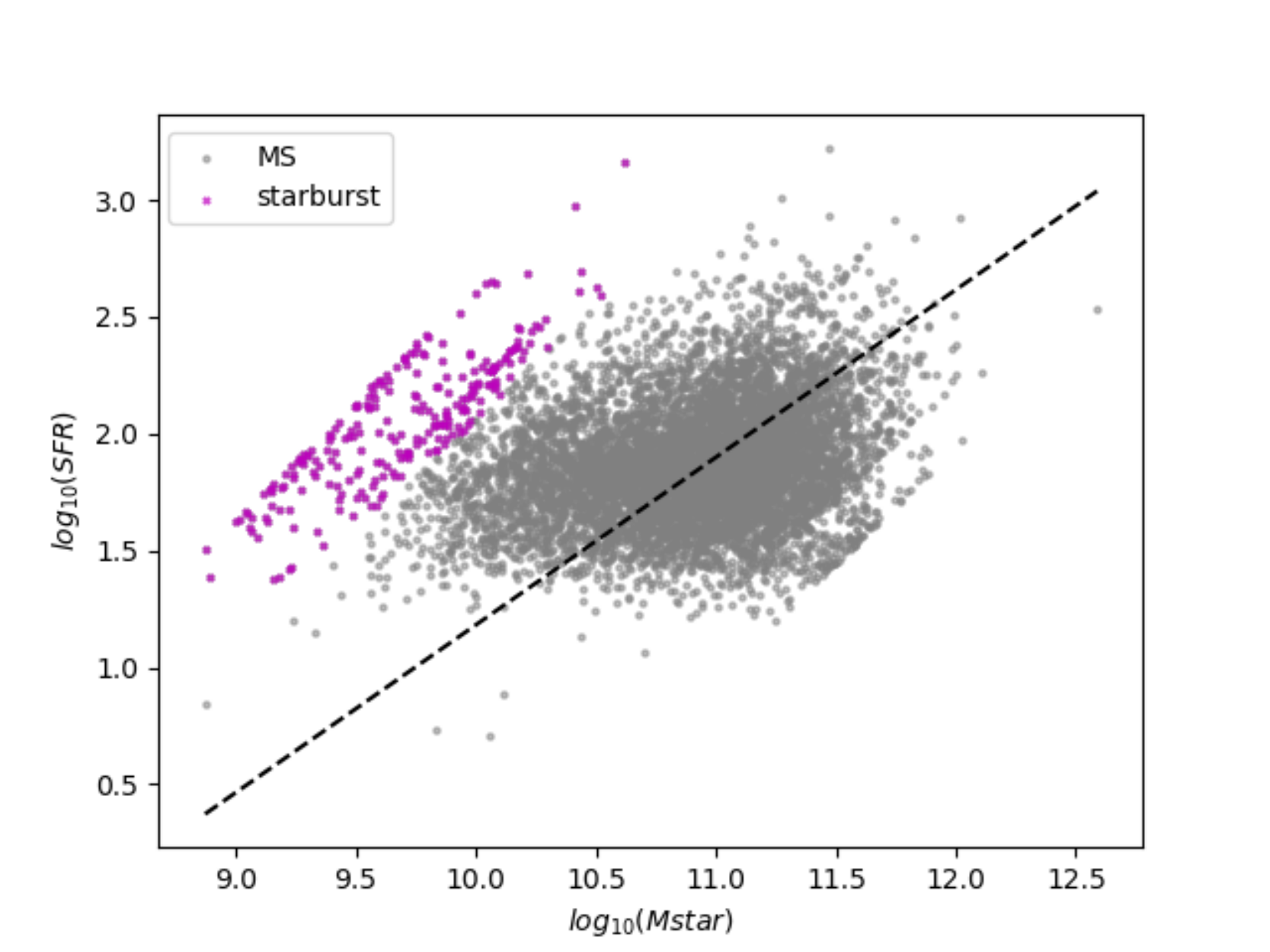}
    \caption{\textit{left panel}: Example of the sSFR distribution of \EN galaxies in the redshift range $1<z<1.5$, obtained with delayed SFH plus additional burst. The orange line respresents the Gaussian fit, the black dashed line corresponds to the division between starburst (SB galaxies are located on the right side of the line) and MS galaxies. The division is located $3\sigma$ away from the Gaussian's centre. \textit{Right panel}: The MS distribution for the same redshift bin. Magenta circles represent SB galaxies selected from the sSFR distribution shown in the left panel, the black dashed line represents the \citealt{Speagle2014} MS.}
    \label{fig:sSFR}
\end{figure*}

To interpret possible bias for the physical parameter estimation we have to ensure that the  selection effects do not produce artificial trends in the analysis.  
To quantify the accuracy of the physical properties estimates of LSST galaxies we decide to focus on MS objects only, as we can select a large number of those galaxies from the HELP data to obtain a statistically important sample of real and simulated galaxies.  
The method we used to separate MS galaxies is described in detail in \cite{Rodighiero2011}. 
We divided our sample in four redshift bins (Table~\ref{table:SB percentage}), as the definition of starbursts changes with redshift. 
In fact, it was shown by \cite{Schreiber2015} that the average SFR of star forming galaxies, in the same ranges of masses, increases with redshift. In this work starbursts are defined according to their specific SFR distribution (SFR/\mstar, hereafter sSFR). 
Figure~\ref{fig:sSFR} shows that the sSFR follows a Gaussian distribution. 
We follow the same definition of starbursts as \cite{Rodighiero2011}, i.e objects with sSFR that lie above $\overline{\rm{sSFR}}$+$3\sigma$, where $\overline{\rm{sSFR}}$ is the Gaussian mean of the sSFR distribution.
The right panel in Fig.~\ref{fig:sSFR} shows the selected starbursts and the MS galaxies.

\begin{figure*}[htb]
    \centering
    \includegraphics[width=0.8\hsize]{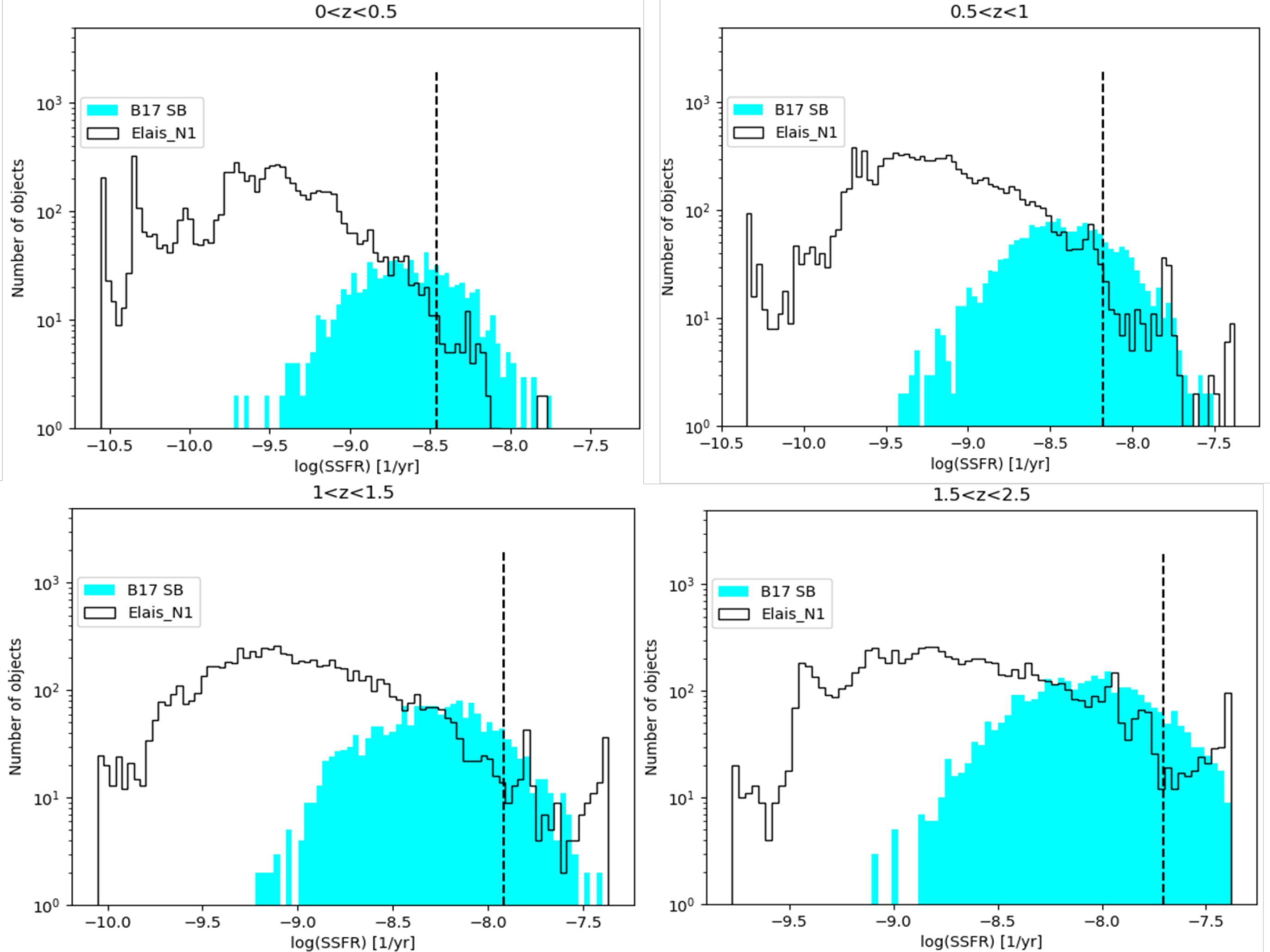}
    \caption{The sSFR distribution in four different redshift bins, obtained with delayed SFH plus additional burst. Open histograms are the galaxy distributions (MS+SB) derived in this work, the black dashed line is the division between SBs (right side of the line) and MS (left side of the line) in our sample, while cyan full histograms represent simulated SB sample from \citetalias{Bethermin2017}. The division is located around  $3\sigma$ from the Gaussian centre. The sample is the same as for Fig. \ref{fig:sSFR}, but the binning is different. }
    \label{fig:sSFR_comparison}
\end{figure*}

To further test the reliability of the SB selection we compare our distribution and the position of starbursts  with the one found in \cite{Bethermin2017} (hereafter: \citetalias{Bethermin2017}), a catalogue of simulated galaxies. The \citetalias{Bethermin2017} is built on IR/sub-mm data and it is one of few models that are able to simultaneously match the total IR number counts and the evolution of sSFR. It simulates 2~$deg^2$ field including physical clustering from dark matter simulation, and is thus perfectly suitable for the comparison purpose.
Figure~\ref{fig:sSFR_comparison} shows the comparison between SB distribution derived in this paper and the sample of simulated SBs from \citetalias{Bethermin2017} (cyan distribution). 
The simulated SB sample extends to sSFR values lower than the sSFR range obtained from our analysis. 
The discrepancy can be explained by the use of two different selection methods in our work and in \citetalias{Bethermin2017}. On the one hand, \citetalias{Bethermin2017} randomly draw the SFR of each source using a continuous log-normal distribution \citep[in agreement with the
observational results, e.g. ][]{Rodighiero2011} and then used the \cite{Schreiber2015} definition of the MS to select the galaxies, with an additional offset correction for galaxies at z$<$0.5. Specifically, \citetalias{Bethermin2017} defined MS objects as those belonging to the distribution centred on 0.87$\times \rm{SFR_{MS}}$ with 0.2 dex of width, and SBs as the ones belonging to the distribution centred at 5.3$\times\rm{SFR_{MS}}$ with 0.3 dex of width. 
On the other hand, our selection is based on the statistical analysis presented in \cite{Rodighiero2011} which uses the sSFR distribution over a broad redshift range, which makes our selection more discrete while \citetalias{Bethermin2017} selection is continuous. Figure~\ref{fig:sSFR_comparison} shows that the purity of our selection is very high but, at the same time, it can be incomplete for less active galaxies.

In total we remove 763 SB galaxies from \EN and 228 from COSMOS. Table~\ref{table:SB percentage} shows the fraction of SB galaxies found in the \EN and COSMOS field. 
Our findings for the \EN field agree with the literature, as it is expected that the fraction of SBs rises from 1\% at low redshift to around 3\% at higher redshift, and remains flat thereafter (i.e. \citealt{Bethermin2012a}).

\begin{table}
\caption{Total number of galaxies and SB percentage for ELAIS and COSMOS fields, in each redshift range.}   
\label{table:SB percentage}      
\centering                          
\begin{tabular}{c c c c c}        
\hline\hline                                       
Redshift range & \EN & SB\% & COSMOS & SB\% \\\\ 
\hline
0 - 0.5 & 7\,718 & 0.64\% & 2\,629 & 1.35\% \\
0.5 - 1 & 11\,353 & 1.54\% & 5\,081 & 1.65\% \\
1 - 1.5 & 8\,120 & 2.96\% & 2\,526 & 2.09\% \\
1.5 - 2.5 & 9\,214 & 3.19\% & 3\,733 & 0.83\% \\
\hline
\end{tabular}
\vspace{1mm}
\end{table}

\subsection{Passive galaxies selection }
\label{subsec:Passive}
The SFH with two or more stellar populations is suitable to fit active galaxies, with moderate or high SFR. 
For this reason we analyse the remaining objects without SBs employing the delayed SFH in the SED fitting, which is more suitable for normal MS galaxies \citep{Ciesla2016}.
For the simplicity of the  analysis, after removing SB galaxies, we replace delayed SFH with additional burst  by simple delayed SFH. 
All the parameters used for the SED fitting are listed in Tab.~\ref{table:appendix parameters}. We performed again the SED fitting to obtain real physical properties of the sample of real MS galaxies.
Those values were next used to simulate the LSST observations (see Section~\ref{Subsec:simulatedLSSTdata}).

To ensure of the purity of the MS sample we additionally removed possible passive galaxies.
As for the starburst evaluation, many different methods were employed in the literature to select red passive galaxies, i. e. UVJ and NUVrK colour diagram analysis (\citealt{Williams2009}, \citealt{Arnouts2013}), division based on sSFR (\citealt{Vulcani2015}, \citealt{Salim2016}, \citealt{Salim2018}), or  unsupervised machine learning (\citealt{Gosia2018}). We decide to follow \cite{Salim2018} method by removing all objects with $ log_{10}\rm(sSFR [yr^{-1}])<-11\mbox{ }$. In this way we remove 340 (1\%) and 63 (0.5\%) galaxies from \EN and COSMOS field respectively.
The almost negligible number of passive galaxies in the HELP sample is  related to our initial sample selection, which required at least two \Herschel measurements with S/N$>$3. 

\subsection{AGN contribution}
\label{subsec:AGNs}

Taking advantage of IRAC detection for all galaxies included in our analysis we used MIR detections to find how numerous is the AGN population in our sample. We employed two different selection criteria, based on MIR photometry (IRAC bands) analysis, explained in detail in \cite{Stern2005} and \cite{Donley2012}. Fig. \ref{fig:appendixAGN} shows the IRAC colour-colour selection using \cite{Donley2012} (upper panels) and \cite{Stern2005} (lower panels) methods. Using both criteria, we find a negligible number of AGN, in comparison to the final sample (1.56\% and 5.16\% for \citealt{Donley2012} and \citealp{Stern2005} criterion, respectively, see Table~\ref{table:appendix} for detailed information for both fields).
The redshift distribution of  selected AGNs is shown in Fig.~\ref{fig:appendixz}.  
For consistency with the cuts made previously we removed AGNs from our sample. 
We decide to use a conservative approach, and we removed all 2\,603 possible AGNs found with the \cite{Stern2005} method, as this selection includes all the AGNs detected with the \cite{Donley2012} approach.

\begin{table*}
\caption{Number of AGN selected based on the MIR features for \EN and  COSMOS field. The last column shows the total number (and percentage) of AGNs in the full sample.}    
\label{table:appendix}      
\centering                          
\begin{tabular}{c c c c }        
\hline\hline                                       
Method & \EN & COSMOS & Total sample  \\\\ 
\hline
\cite{Stern2005} & 1\,269~(3.48\%) & 1\,334~(9.50\%) & 2\,603~(5.16\%)\\
\cite{Donley2012} & 497~(1.36\%) & 291~(2.08\%) & 788~(1.56\%)\\
\hline
\end{tabular}
\vspace{1mm}
\end{table*}

\subsection{Outlier selection}
\label{subsec:outlier}
Due to the large and unknown number of galaxy's free parameters, a simple $\chi^2_r$ selection cannot assure us to remove the majority of the outliers from our sample. 
In order to eliminate possible outliers and to assure the high quality of the SED fitting, along with a $\chi^2_r$ selection, we use an estimation of physical properties $L_{dust}$ and $M_{star}$ (See appendix \ref{app:outlier}). 
A similar procedure was used by  \cite{Malek2018} for HELP \EN field. Based on that criteria  we removed 2\,117 galaxies from \EN, and 640 from COSMOS field, 5.81\% and 4.75\%, respectively.

\subsection{Final sample}

\begin{figure}[htb]
    \centering
    \includegraphics[width=\hsize]{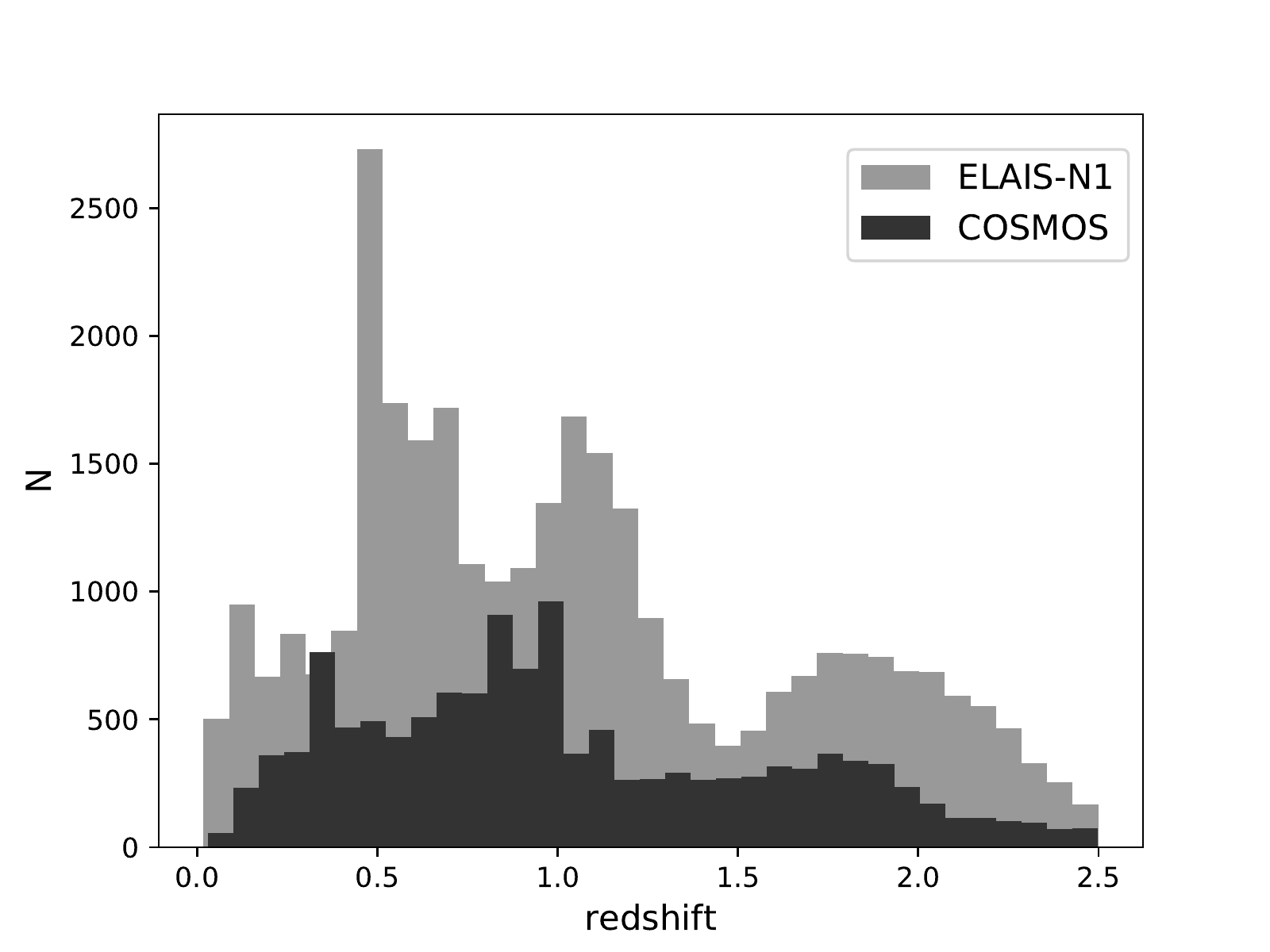}
    \caption{Redshift distribution of ELAIS-N1 and COSMOS samples based on \cite{Duncan2018} approach.}
    \label{fig:Zdistr}
\end{figure}

To obtain the final sample of normal star-forming galaxies, we removed possible starbursts (Sec.~\ref{subsec:Starburst}), passive galaxies (Sec.~\ref{subsec:Passive}), and possible AGNs (Sec.~\ref{subsec:AGNs}). We also did additional cleaning using outlier selection (Sec.~\ref{subsec:outlier}) to remove all galaxies with possible wrong photometry, or wrong matches between UV-optical and FIR measurements. 
At the end of the process we remain with 31\,936 objects for \EN (87\% of the total sample) and 11\,716 galaxies for COSMOS (84\% of the total number). Furthermore, in order to validate the photometric redshift estimates used for this objects, we perform a comparison with spectroscopic redshift estimates, available for $\sim$ 5000 galaxies in \EN and COSMOS fields. Following \cite{Duncan2018} definition of critical outlier ($\frac{|\Delta z|}{1+z_s}>0.2$), we find the fraction of outlier in our sample at the level of 4\%, in agreement with what found in previous works (\citealt{Ilbert2009}, \citealt{Hildebrandt2010}, \citealt{Duncan2018}). The final redshift distributions of both samples are shown in Fig.~\ref{fig:Zdistr}.


\section{LSST physical properties estimation}
\label{Sec:Results}
In the following section we discuss the LSST data and uncertainties from simulations and the estimation of the physical properties of the galaxies obtained by performing the SED fitting of: i) the fiducial input parameters plus LSST data only; ii) the fiducial input parameters plus LSST data coupled with other observations.

\subsection{LSST simulated data and uncertainties estimation}
\label{Subsec:simulatedLSSTdata}

LSST data simulation has been a popular topic in the last years, considering the upcoming start of the survey \citep{Ivezic2019}. 
In this work we derive an 'LSST-like catalogue' from the best fit of the observational data, described in Sect.~\ref{Sec:Data}. In this way, we are able to quantify the difference between the estimation of the physical properties based on the LSST measurements only and the UV-to-FIR wavelength of the real, observed  objects. 
Considering the 
depth reached with 10 years worth of survey data, it is very likely that LSST will observe objects not visible with the current ground survey telescopes, and this work will be a starting point to learn how to treat those objects with SED fitting methods.

We simulate the observed fluxes in the six LSST bands \citep[\textit{ugrizy}, of which the filter response curve is provided by the LSST developers team,][]{Ivezic2019}. 
To obtain LSST fluxes we run CIGALE by fitting the  photometric measurements and providing to the code the LSST filter response curves. 
We use a CIGALE module (called \textit{fluxes}) specifically designed to estimate the fluxes in the defined filters. 
We compute LSST fluxes from the best-fit model of each object.
We include in our sample all the galaxies that will be detected in all bands at the depth of the 10 year survey: $u<$26.1, $g<$27.4, $r<$27.5, $i<$26.8, $z<$26.1, $y<$24.9. In this way we discard 8\,645 objects (23\% of the total sample) from \EN and 2\,623 (19\% of the total sample) from COSMOS.

To incorporate an LSST-like observational uncertainty into our catalogue, we must take into account random phenomena that could occur during a real observation, such as change in the sky seeing, number of visits etc.
The predicted magnitude errors, that we convert in flux errors following the conversion provided in the LSST manual \citep{Ivezic2019}, depend on the galaxy's magnitude, the sky seeing, and the total survey  exposure time in a given filter. 
We use the LSST simulation software package \textit{CatSim} \footnote{\url{https://www.lsst.org/scientists/simulations/catsim}} to calculate magnitude errors. 
The aforementioned error evaluation is based on Eq.~5 of \cite{Ivezic2019}, and takes into account variations in the photometry due to hardware and observational components (e.g. detector, darksky, atmosphere). 
The random error evaluated is then divided by the square root of the number of visits during the survey. 
The LSST manual provides mean values for all these components. 

To mimic the real conditions we add to the average value of each component provided in the LSST manual, a value randomly chosen from a Gaussian distribution centred on the provided mean value, having standard deviation the 10\% of the mean.
In this process, we vary the number of visits and sky seeing. The assumed standard conditions for these components are 0.8 arcsec for the seeing, and a uniform progression that assumes a total of 56, 80, 184, 184, 160 and 160 visits in filters \textit{ugrizy} respectively, in 10 years, where each visit is 30 seconds of integration time. 

\begin{figure}[htb]
    \centering
    \includegraphics[width=\hsize]{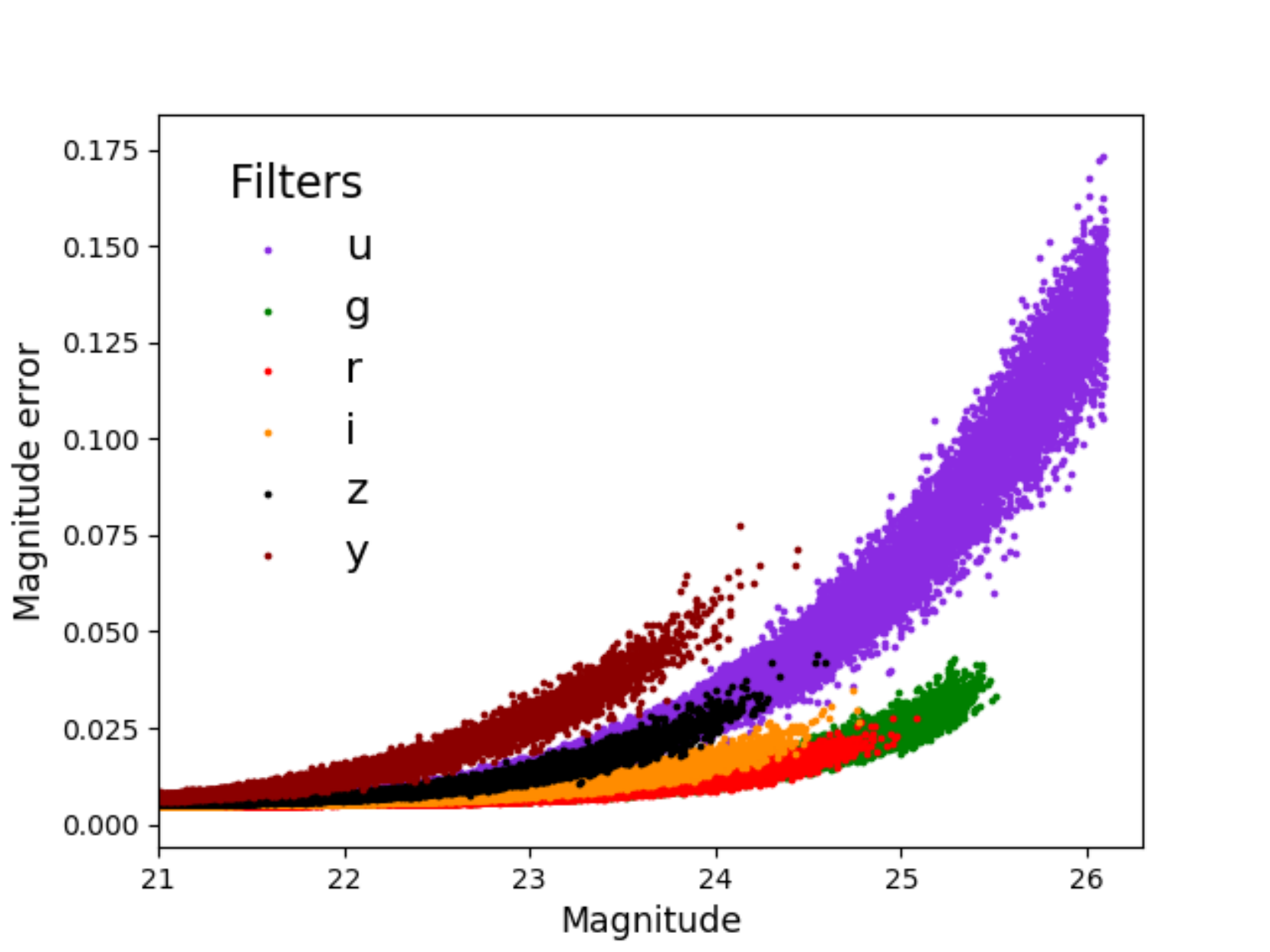}
    \caption{Magnitude errors vs observed apparent magnitudes for simulated LSST observations from the \EN sample. Simulations are cut to the maximum LSST limit. The choice to vary simulation parameters allows us to obtain different errors for similar magnitude values, as we would expect in the case of real observations. The brightest end reached in the paper is 13 mag but for clarity of the plot, we cut it at 21 mag.}
    \label{fig:magvserr}
\end{figure}

At the end, to further mimic a possible divergence of the 'real' observed flux from the simulated value evaluated from the best-fit SED, we again add to the latter a value randomly chosen from a Gaussian distribution centred on 0, having standard deviation the flux error calculated before.
Figure~\ref{fig:magvserr} shows the magnitude errors as a function of the simulated observed magnitude for our sample of galaxies. 
We only select objects that would be observed in all six bands according to our simulation. 
As a result, we reach the LSST magnitude limit only for the \textit{u} band. 
The final catalogue contains simulated LSST fluxes and uncertainties for 23\,291 galaxies in the \EN field and 9\,093 in the COSMOS field.  
The catalogues, together with the photometric redshifts and HELP IDs, are available on the HELP virtual observatory\footnote{\url{https://www.herschel-vos.phys.susx.ac.uk}}. 

\subsection{Fiducial parameters and LSST data only}

\begin{figure}
    \centering
    \includegraphics[width=0.9\hsize]{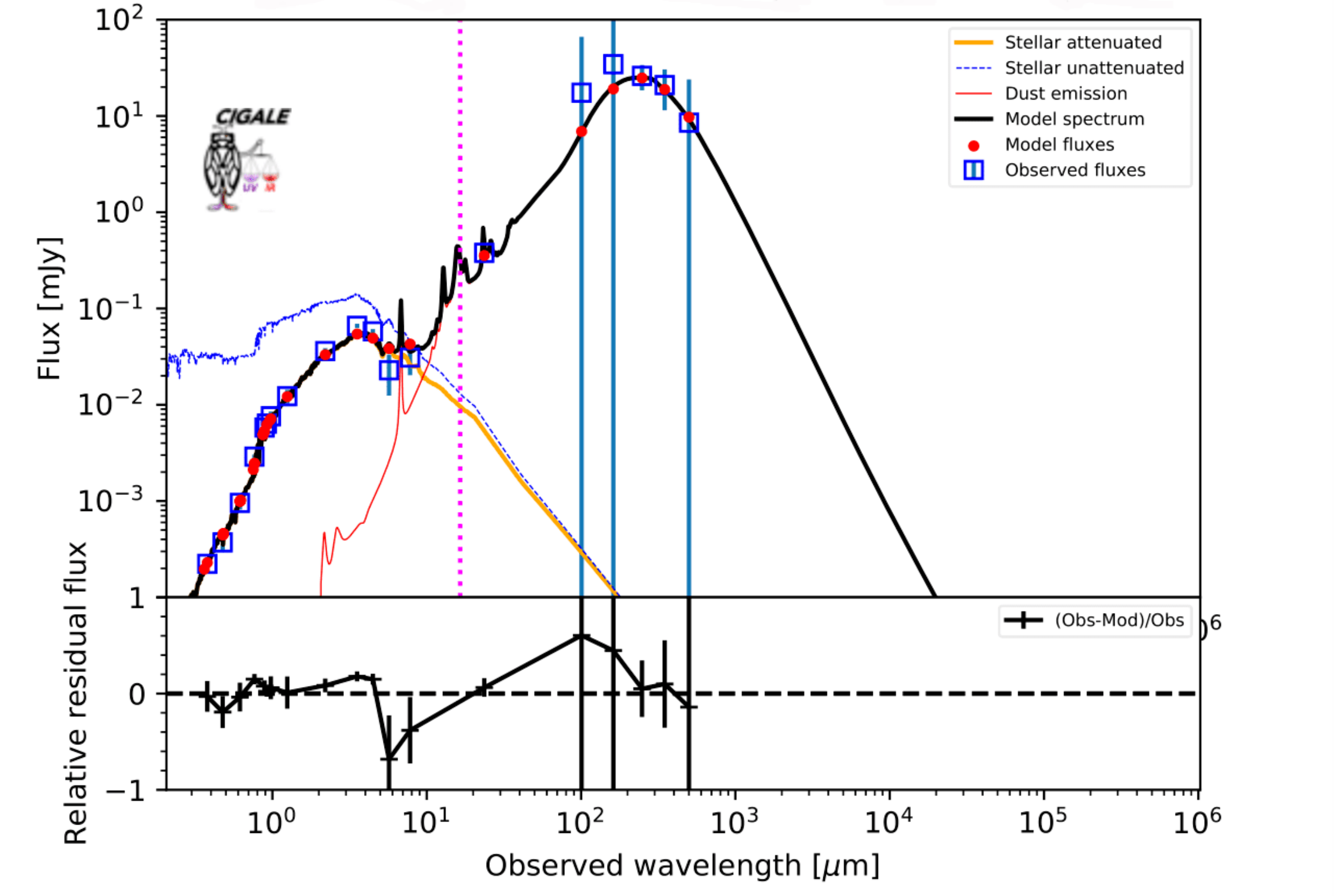}
    \includegraphics[width=0.9\hsize]{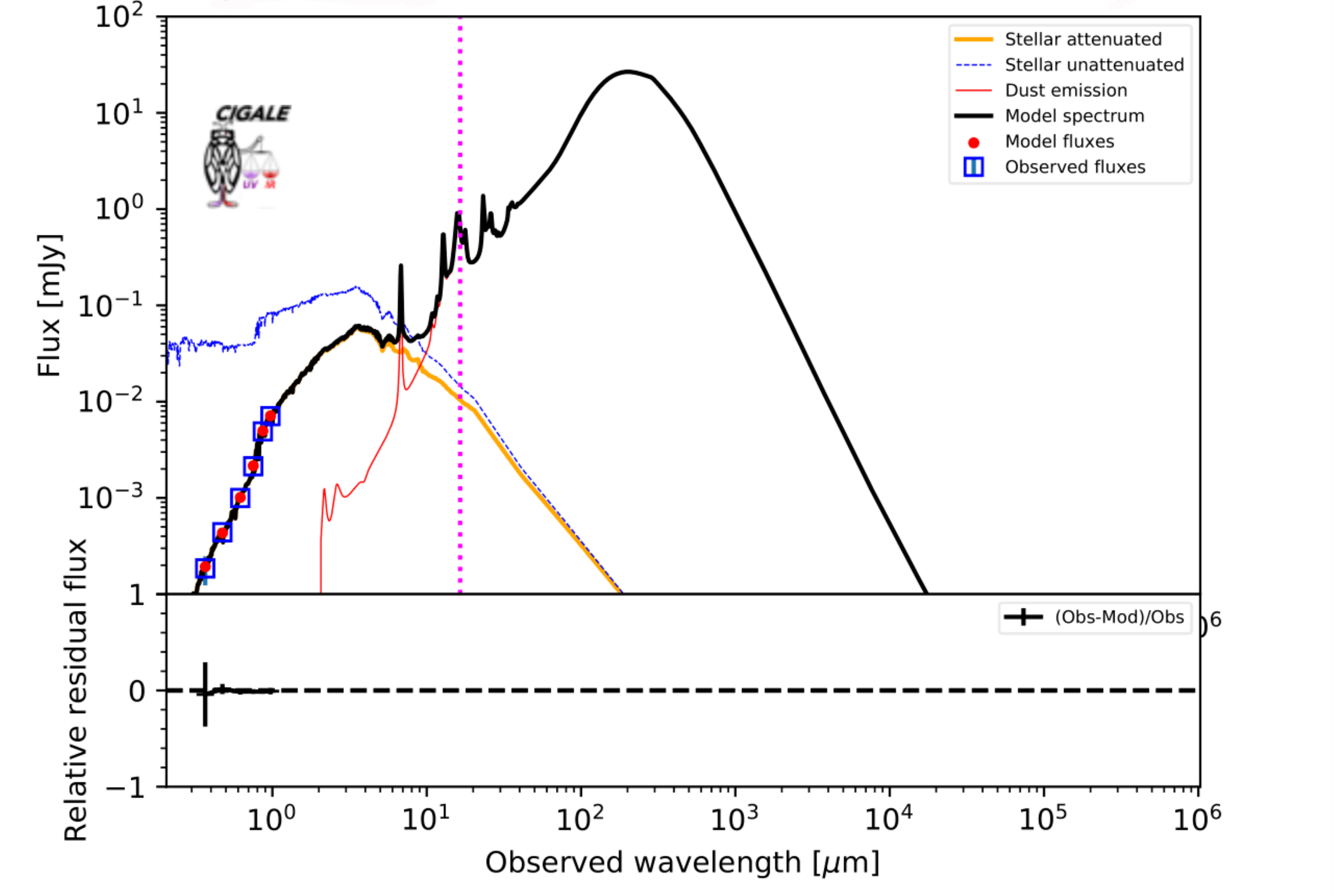}
    \caption{Comparison of SEDs evaluated  for the same object at $z \sim 1$ using the full UV to IR observations (\textit{Upper panel}) and the LSST optical bands only (\textit{Bottom panel}). Blue square represent the observed fluxes, while red dots represent the fluxes predicted by the model.} The magenta lines delimit the division  between optical and IR bands, taken into account by CIGALE. It is possible to observe that the IR emission is not constrained for the LSST estimation.
    \label{fig:SED_example}
\end{figure}

To estimate the main physical properties of the LSST sample, we run CIGALE on simulated LSST observations and uncertainties employing the same modules and parameters used for the HELP MS sample (Table~\ref{table:appendix parameters}, with delayed SFH).
Figure~\ref{fig:SED_example} shows two example SEDs of the same galaxy at redshift 0.92, obtained with the UV-FIR and LSST photometric only data set, respectively. For this specific case, we found an agreement between estimated stellar masses ($M_{star\mbox{ }real}=6.05\cdot10^{10}\pm5.62\cdot10^{9} \mbox{ } M_\odot$, $M_{star\mbox{ }LSST}=5.59\cdot10^{10}\pm1.80\cdot10^{10}\mbox{ } M_\odot$). Instead, the SFR calculated for the LSST-like photometric data only is highly overestimated (by a factor of six) with respect to the real value obtained by employing the UV-FIR data set (SFR$_{real}=11.9\pm2.16 \mbox{ }M_\odot yr^{-1}$ and SFR$_{LSST}=67.4\pm45.9 \mbox{ } M_\odot yr^{-1}$). We can also notice that the residuals for LSST are very small (but never null), as we can easly find a model that almost perfectly fits just 6 observation in the optical part. However, the IR part of the SED, and so the dust emission module, is completely unconstrained (in our work we used the same dust emission module, \cite{Draine2014}, as it was used for the original HELP data with the same grid of parameters). 


\begin{figure*}
    \centering
    \includegraphics[width=0.9\hsize]{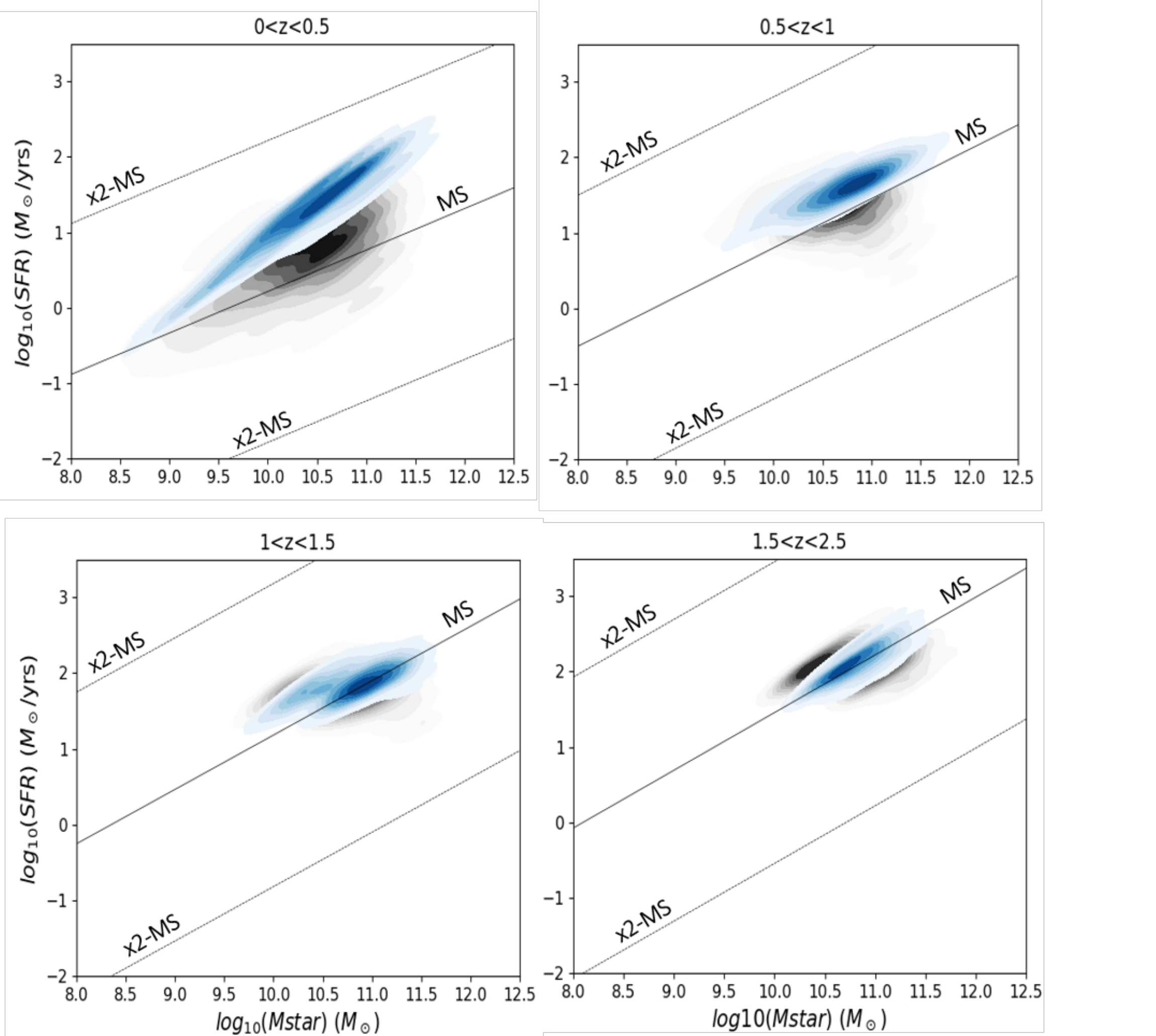}
    \caption{The  main sequence (SFR vs \mstar) relation for the \EN and COSMOS fields in four redshift bins. In blue is represented LSST-like sample, while in black the real sample. The solid black line represent the MS by \citealp{Speagle2014}, while the dashed lines mark the loci 2 times above and below the MS.   
    This plot shows a clear SFR overestimation obtained using LSST bands only, that tends to disappear moving to higher redshift ranges.}
    \label{fig:MS}
\end{figure*}

\begin{figure*}
    \centering
    \includegraphics[width=0.4\hsize]{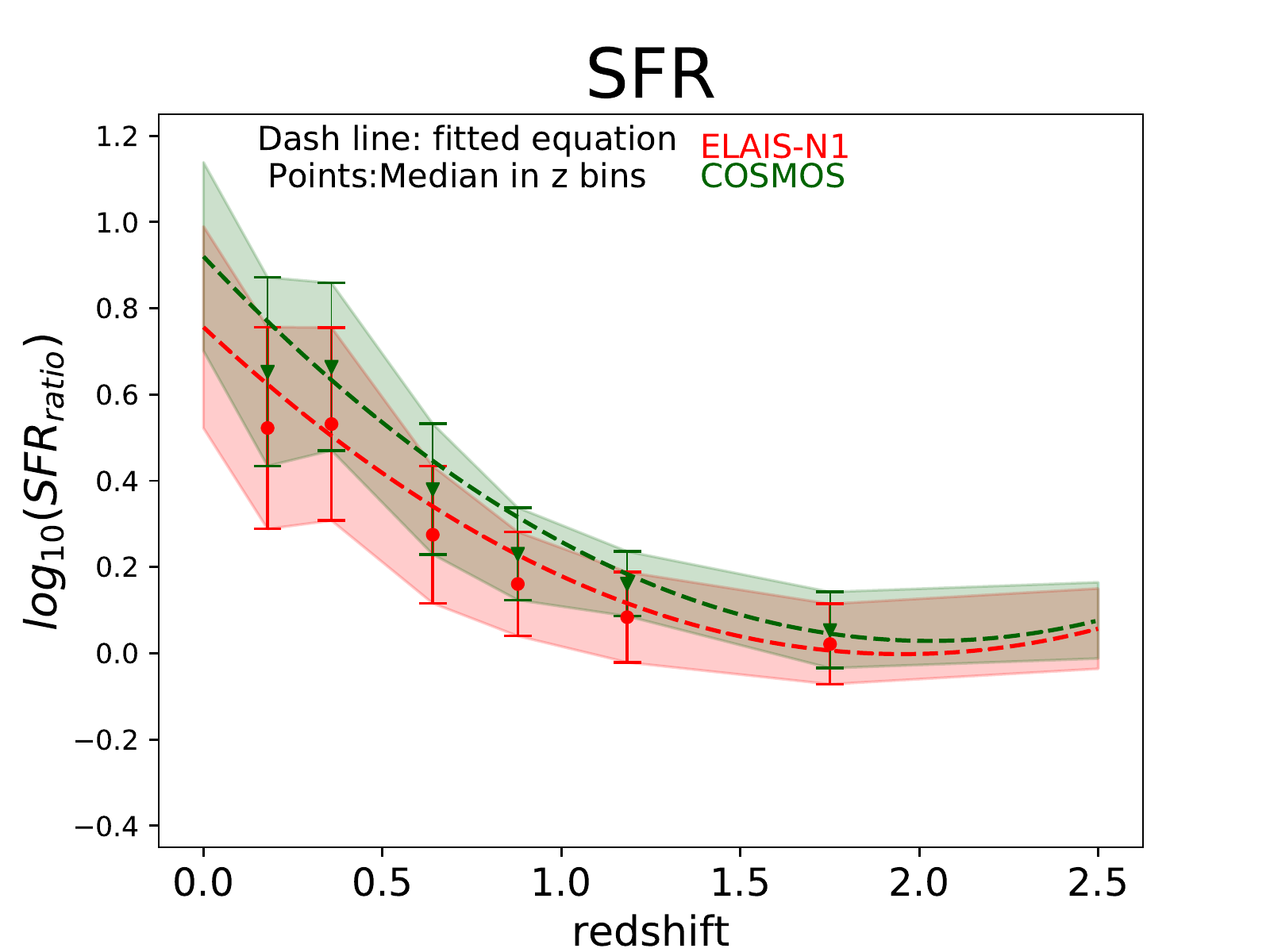}%
    \includegraphics[width=0.4\hsize]{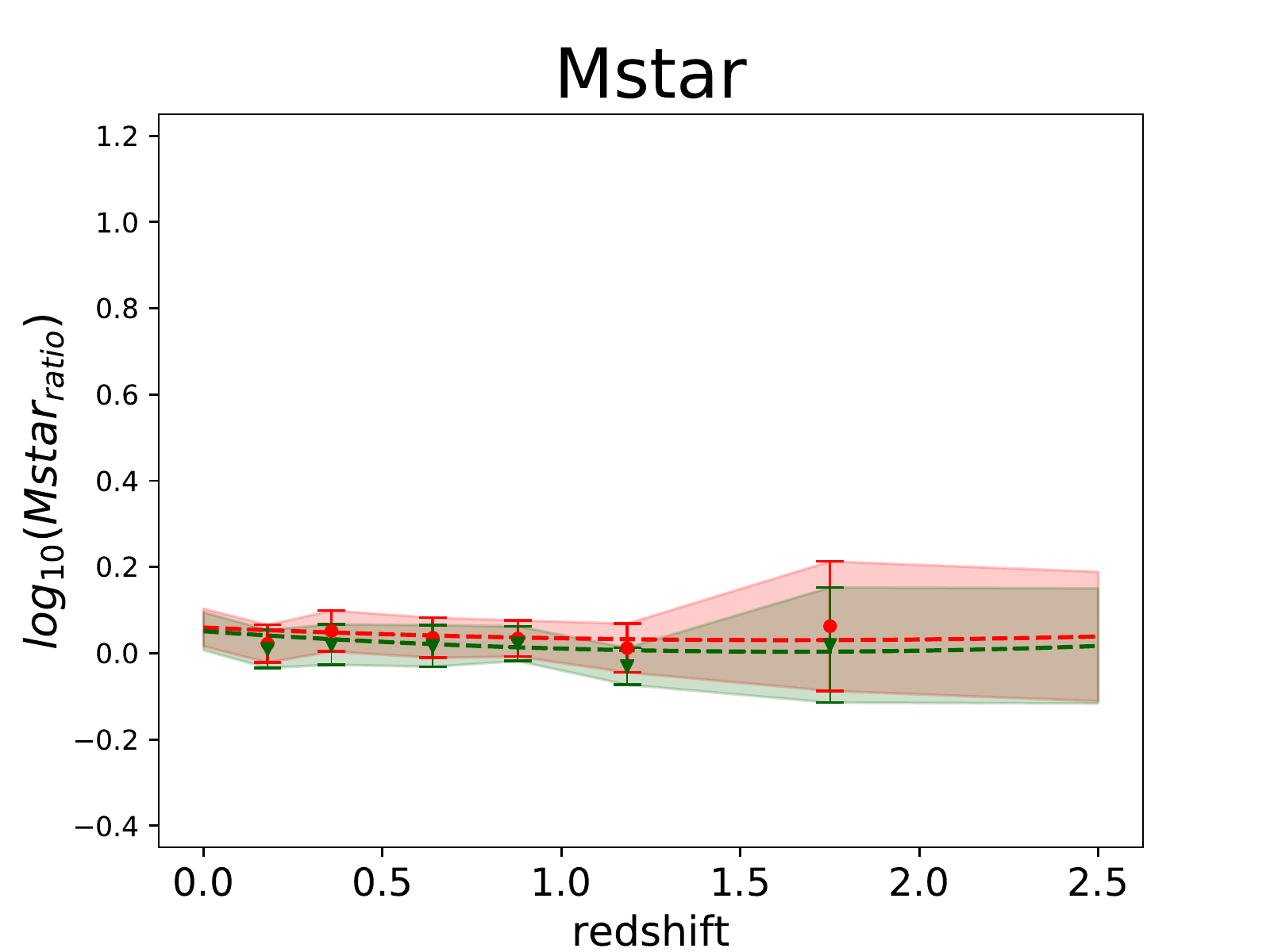}
    \includegraphics[width=0.4\hsize]{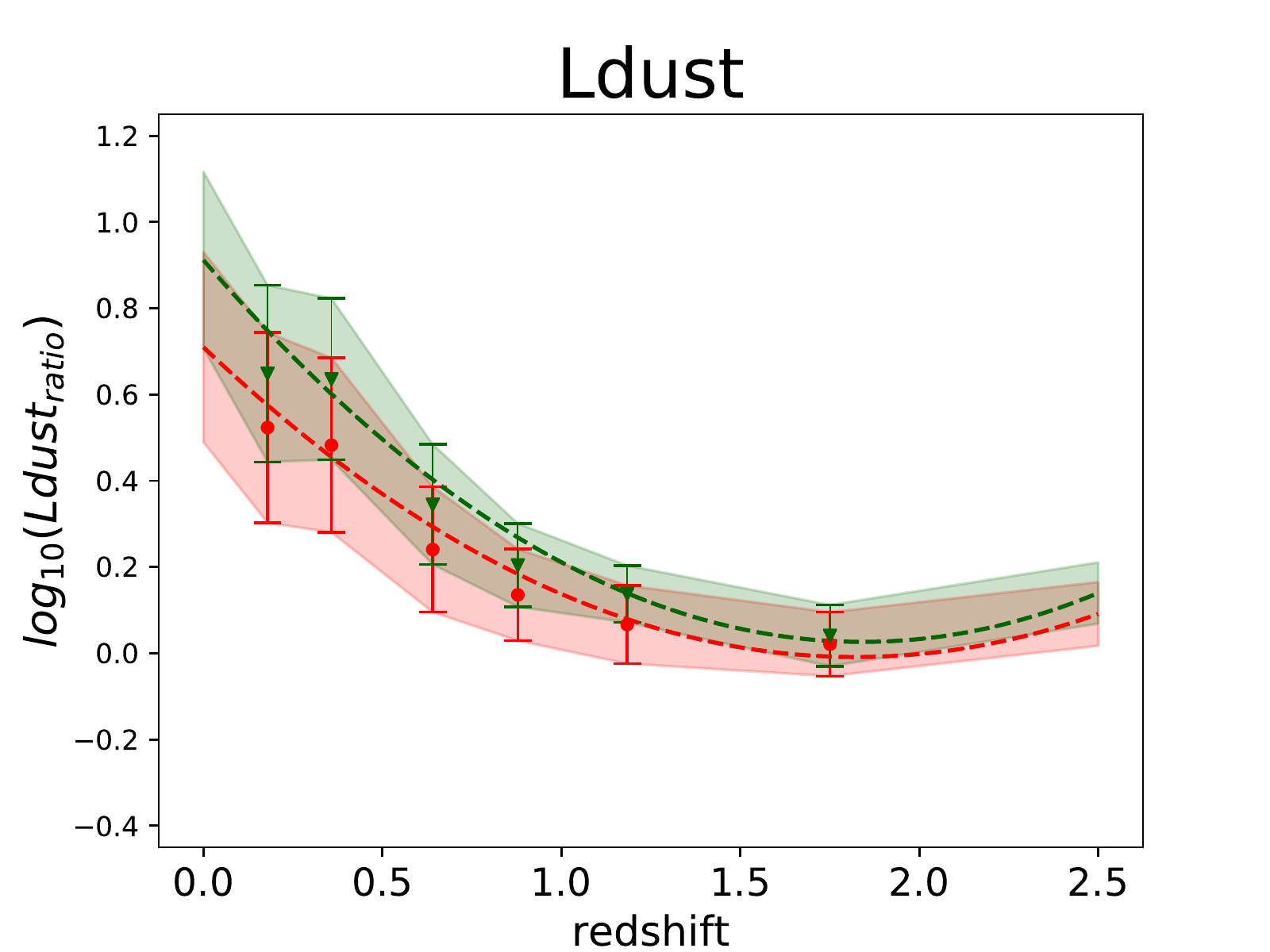}%
    \includegraphics[width=0.4\hsize]{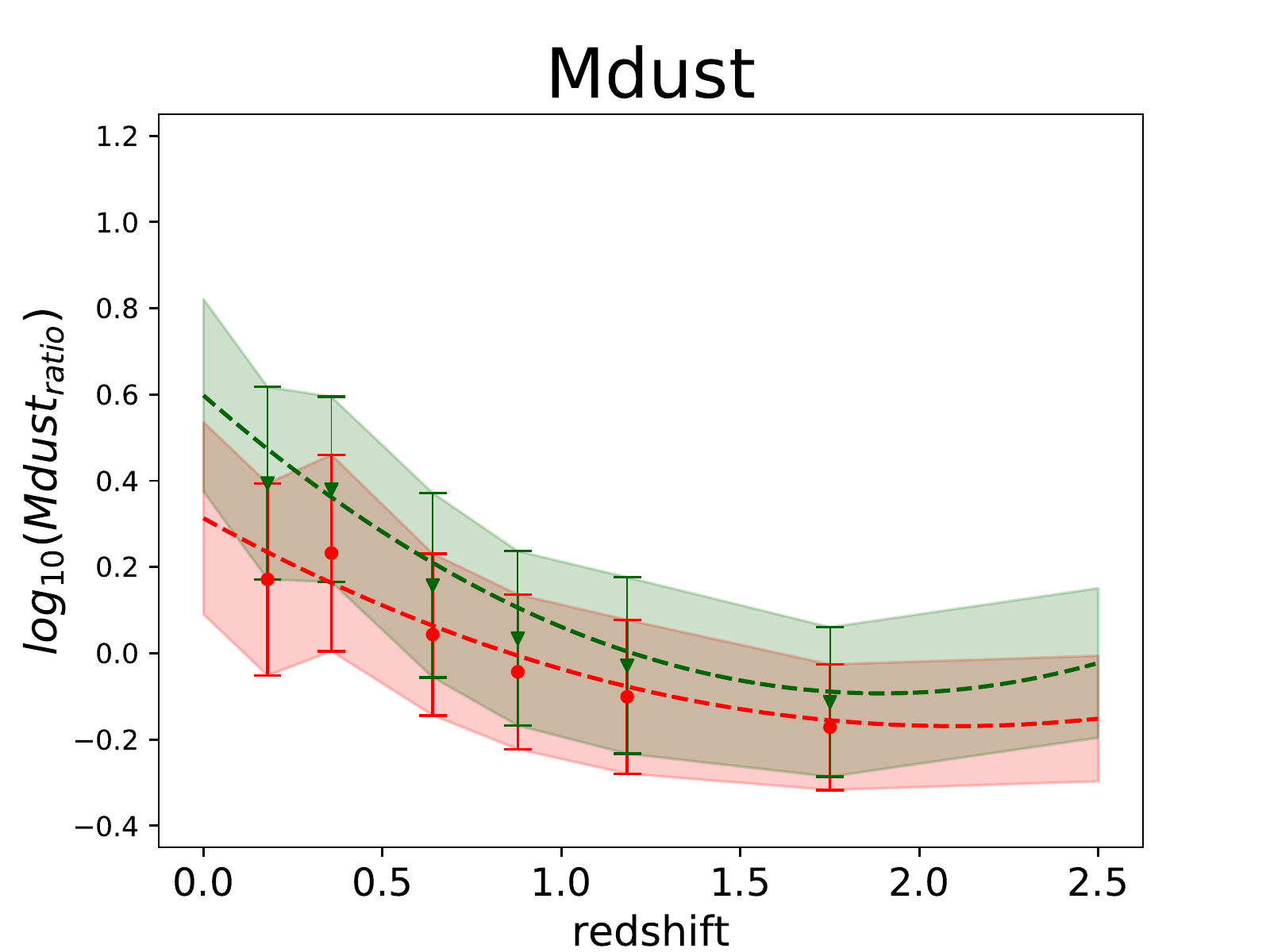}
    \caption{Ratios between different physical properties obtained from the fitting of the simulated LSST data only and from the UV-FIR SED (e.g. $SFR_{ratio}= SFR_{LSST}/SFR_{UV-FIR}$), as a function of the redshift for the \EN and COSMOS fields. The properties obtained from the UV-FIR SED fitting are considered as the true values. From the upper left panel moving clockwise we show: SFR, \mstar, \ldust, and \mdust  comparisons. The dashed lines represent polynomial fits performed on the samples. The points are the median values in each redshift bin, with median absolute deviation as errors. A Ratio equal to zero corresponds to a perfect agreement between estimations obtained based on the LSST-like sample and the real values.
    The distributions calculated for \EN and COSMOS are comparable within the errors. 
     The SFR is systematically overestimated, especially for the lowest redshift range. The stellar masses are instead comparable. At high redshift the stellar mass ratio diverges from zero, mostly because the optical bands are shifted to the NUV range. }
    \label{fig:Density}
\end{figure*}

The relation between SFR and \mstar in the four redshift bins is shown in Fig.~\ref{fig:MS}.
In this Figure, we compare the MS relation obtained for the LSST-like sample with the one obtained from the full UV-FIR SED fitting. We show the MS from \cite{Speagle2014} as a reference for the reader.
We notice that at low redshift the LSST estimation fails to probe low SFR objects, and this leads to a clear division between the respective MS relations, which however overlap at higher redshifts. In Appendix~\ref{app:scatter} we discuss the scatter between our sample and the MS law.   
Figure~\ref{fig:Density} shows this overestimation as a function of redshift, separately for  \EN and COSMOS (\textit{left upper panel}). 
We also plot in the same figure the ratio between the LSST-derived stellar mass, \ldust, and \mdust and the ones from the full UV-IR SED fitting. 
We obtain an overestimation of  the dust related properties (SFR, \ldust, \mdust) while the values of \mstar are comparable.
The overestimation of the SFR is strongly dependent on the redshift. The ratio between the stellar masses is evenly distributed around zero, leading to comparable results between the two runs, as stellar masses mostly rely on optical data. 
This results stands both for \EN and COSMOS fields and shows that there is no dependence on the field.

 The aforementioned results can be explained if we consider how the physical properties are evaluated by the Bayesian method. This is basically done through the likelihood estimation. Each model in the grid of models built from the starting input parameters will have an associated likelihood taken as $exp(-\chi^2/2)$, that is used as weight to estimate the physical parameters (the likelihood-weighted mean of the physical parameters attributed to each model) and the related uncertainty (see Sec. 4.3 of \citealt{Boquien2019}). 
 Fitting just LSST optical observations results in having high likelihood values even for templates that do not reflect the real physical properties of the modelled galaxy. 
 As the SFR is partly estimated from the UV emission of the massive young stars in star forming regions, and 
 this emission is attenuated by the dust and re-emitted in the IR band, we find that the lack of information about the UV and MIR rest-frame wavelengths for the LSST sample results in CIGALE overestimating the attenuation. 
 Overestimated attenuation also results in an overestimation of the SFR.

\begin{figure}
    \centering
    \includegraphics[width=1\hsize]{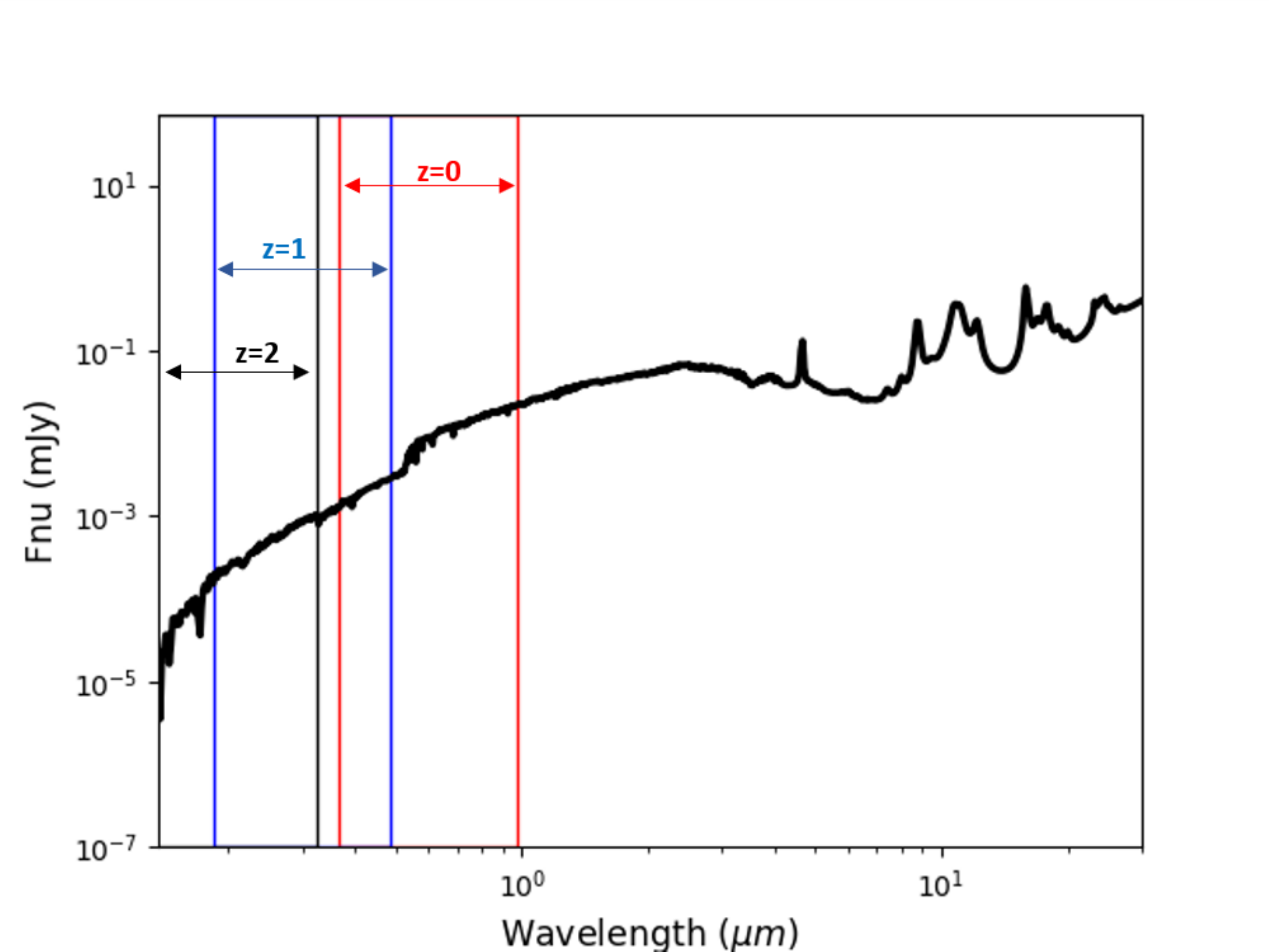}%
    \caption{LSST coverage of an example SED at different redshifts, indicated in the panel. }
    \label{fig:zshift}
\end{figure}

We notice that the fluxes observed by LSST in the optical at 
high redshift are in the UV rest-frame As a consequence, as the UV wavelengths trace young stellar populations, the estimates of the SFR from the SED fitting with CIGALE, significantly improves. Figure~\ref{fig:zshift} shows an example of SED superimposed with LSST coverage at different redshifts. We notice that already at $z=1$, the LSST bands are almost entirely shifted to the rest-frame UV bands, which ranges between 0.01 and 0.38 $\mu m$. As a result, the dust attenuation is better probed since the LSST bands cover a larger portion of the UV rest-frame spectra, where dust attenuates more effectively, and a better constrain of the SFR is provided. As shown in Fig. \ref{fig:Density}, the differences in the estimated SFR become negligible for $z\gtrsim 1.3$.

In order to obtain a useful function to correct the overestimation of the SFR, we perform a polynomial fit on the SFR ratio distribution of \EN and COSMOS combined:
 \begin{equation}
    \rm  \log_{10}({SFR}_{ratio})=0.26\cdot \textit{z}^2 -0.94\cdot \textit{z} +0.87,
 \end{equation}
where  SFR$_{\rm{ratio}}$ stands for SFR$_{LSST}$/SFR$_{real}$. However, this formula is very dependent on the input parameters.

To check if the complex set of parameter used for \cite{Draine2014} model is responsible for the overestimation of the SFR, we run the whole analysis adopting \cite{Dale2014} model. 
Dust emission in this model is parametrised by a single parameter $\alpha$ defined as $\frac{dU}{dM}=U^\alpha$, where M  is the dust mass heated by a radiation field at intensity U.  
We used the parameter $\alpha=2$ as better describe the stellar emission from MS galaxies.
The module allows also to add an optional AGN component, that we set to 0 for this test. 
Comparing the LSST physical properties with the real one using the \cite{Dale2014} module we obtain the same  overestimation of the SFR and \ldust as those obtained with \cite{Draine2014} dust emission. 
We were not able to compare dust masses as it is not an output parameter of the \cite{Dale2014} module. 
As we found no improvement by employing a simpler model, we decided to keep the results obtained with \cite{Draine2014} for the homogeneity with the results published by HELP project \citep[][ Shirley et al., in prep.]{Malek2018}.

\subsection{Fiducial parameters, LSST and ancillary data}

\begin{figure*}
    \centering
    \includegraphics[width=0.5\hsize]{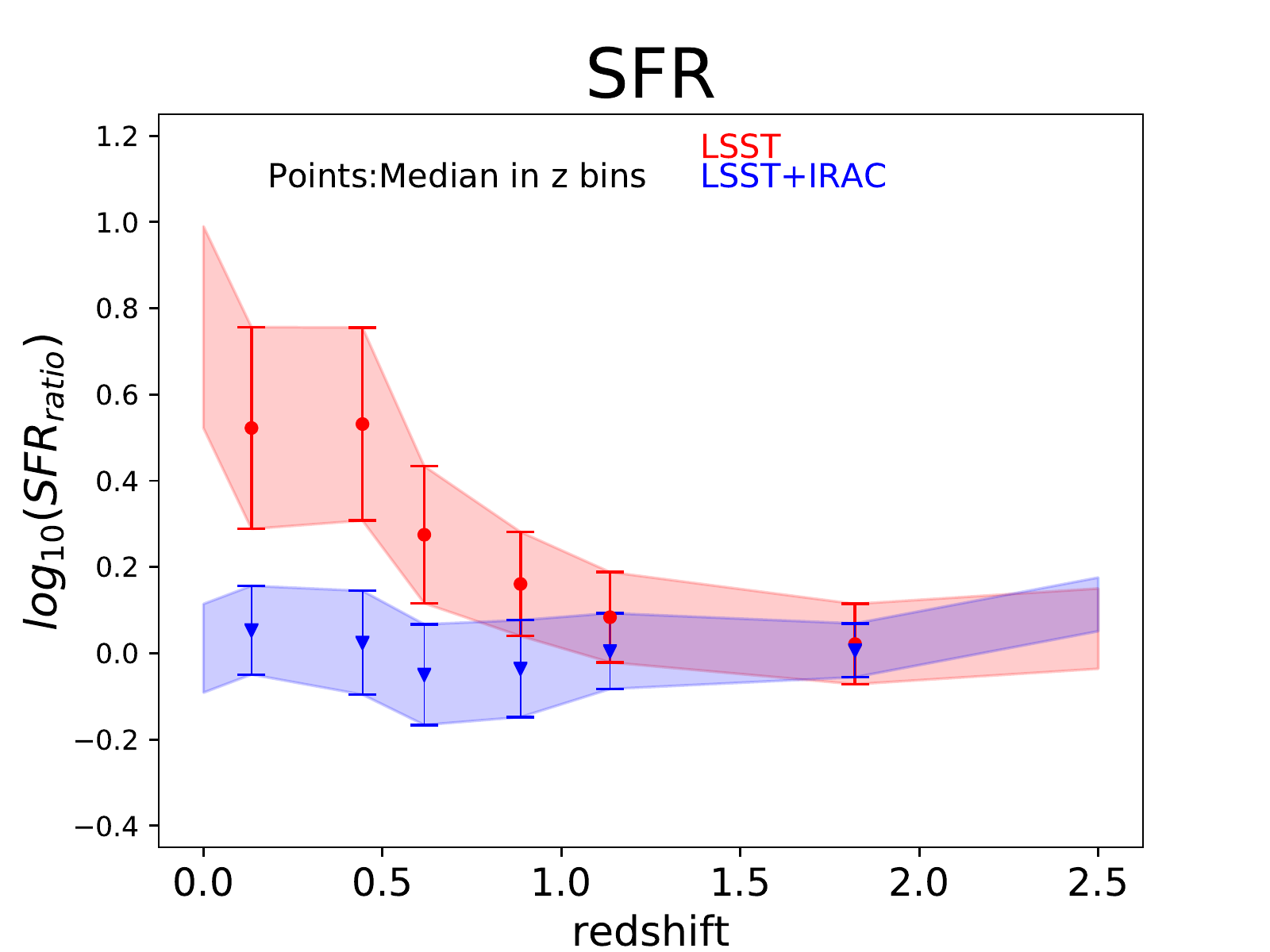}%
    \includegraphics[width=0.5\hsize]{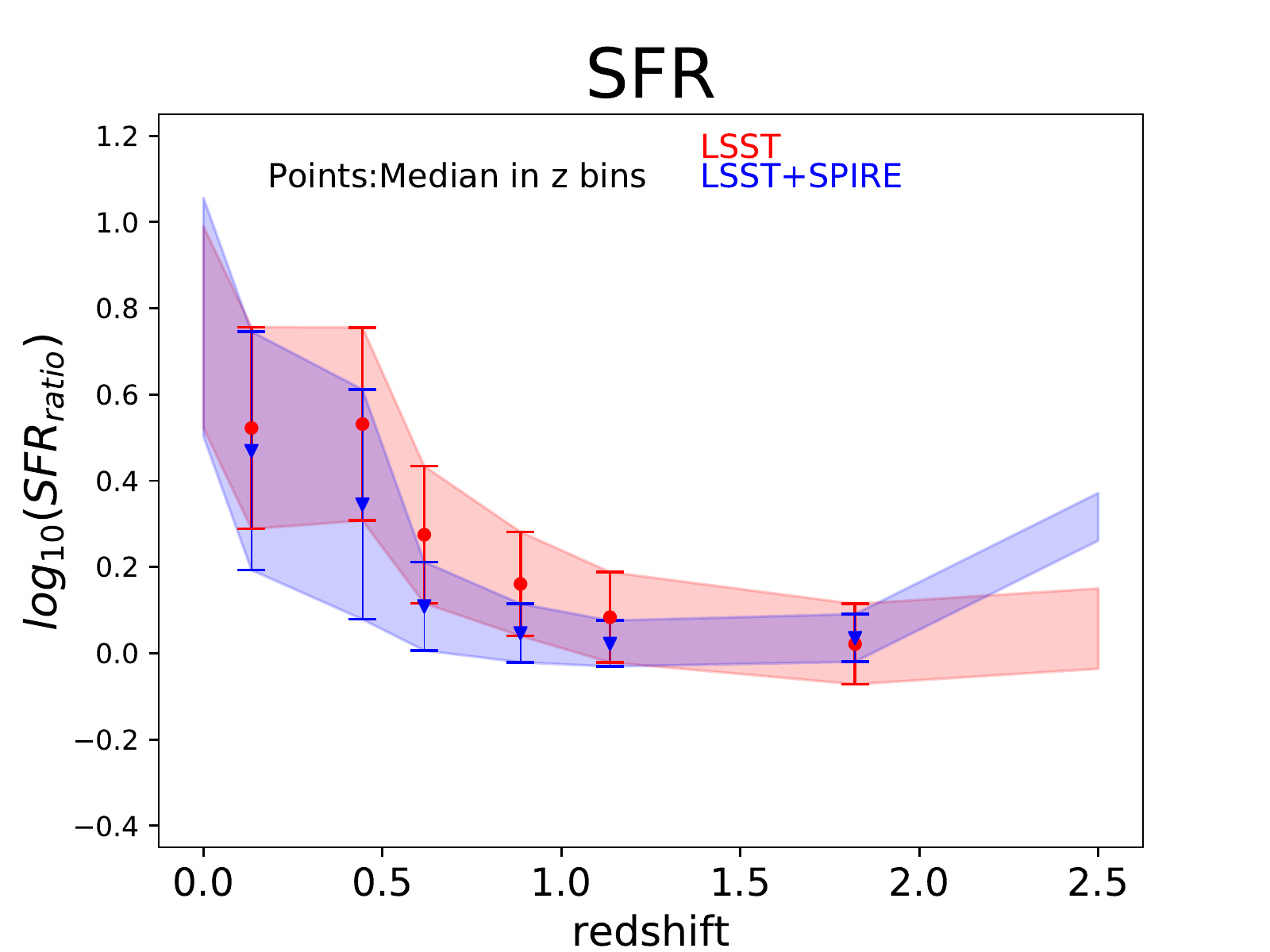}
    \includegraphics[width=0.5\hsize]{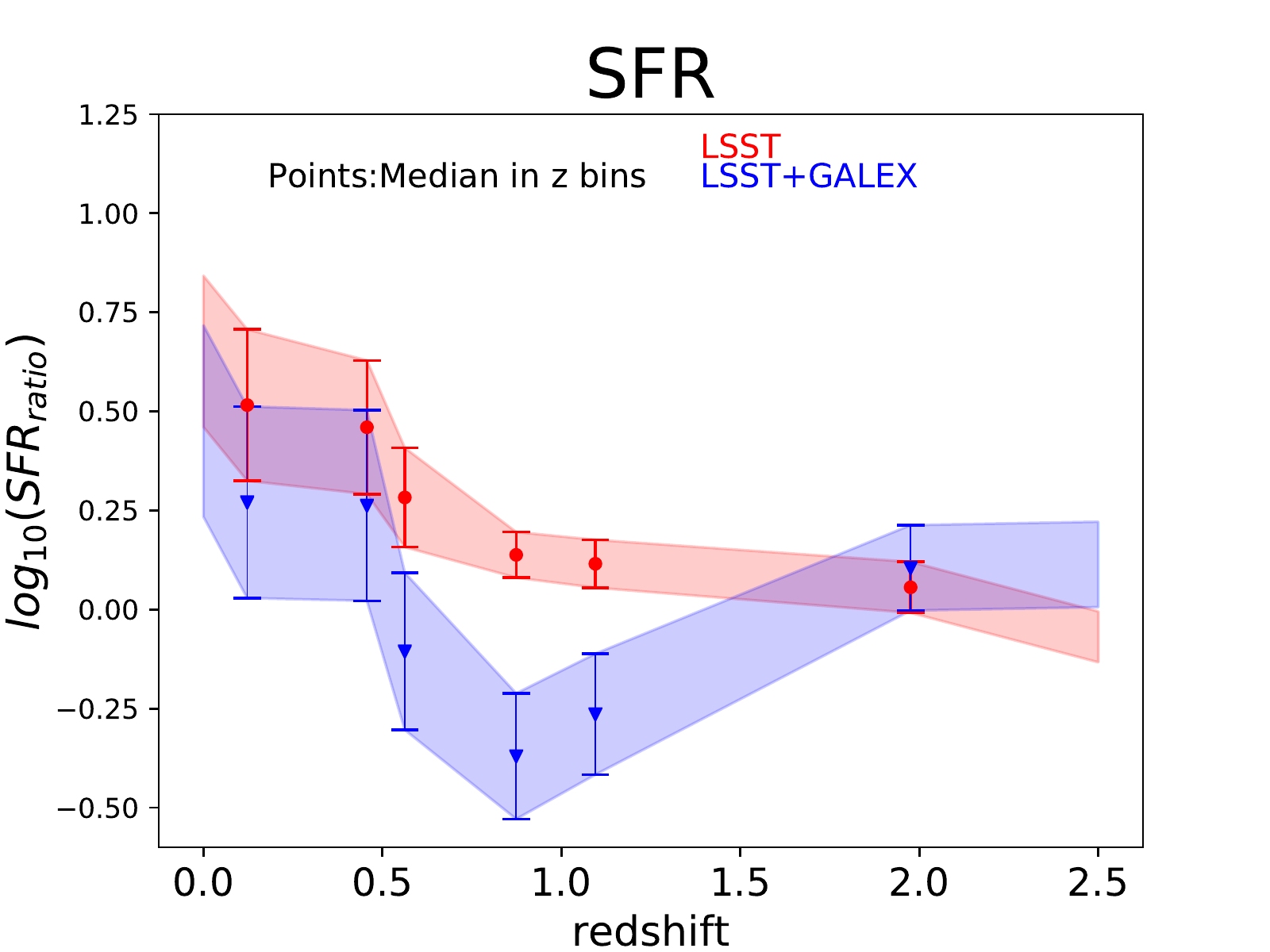}%
    \includegraphics[width=0.5\hsize]{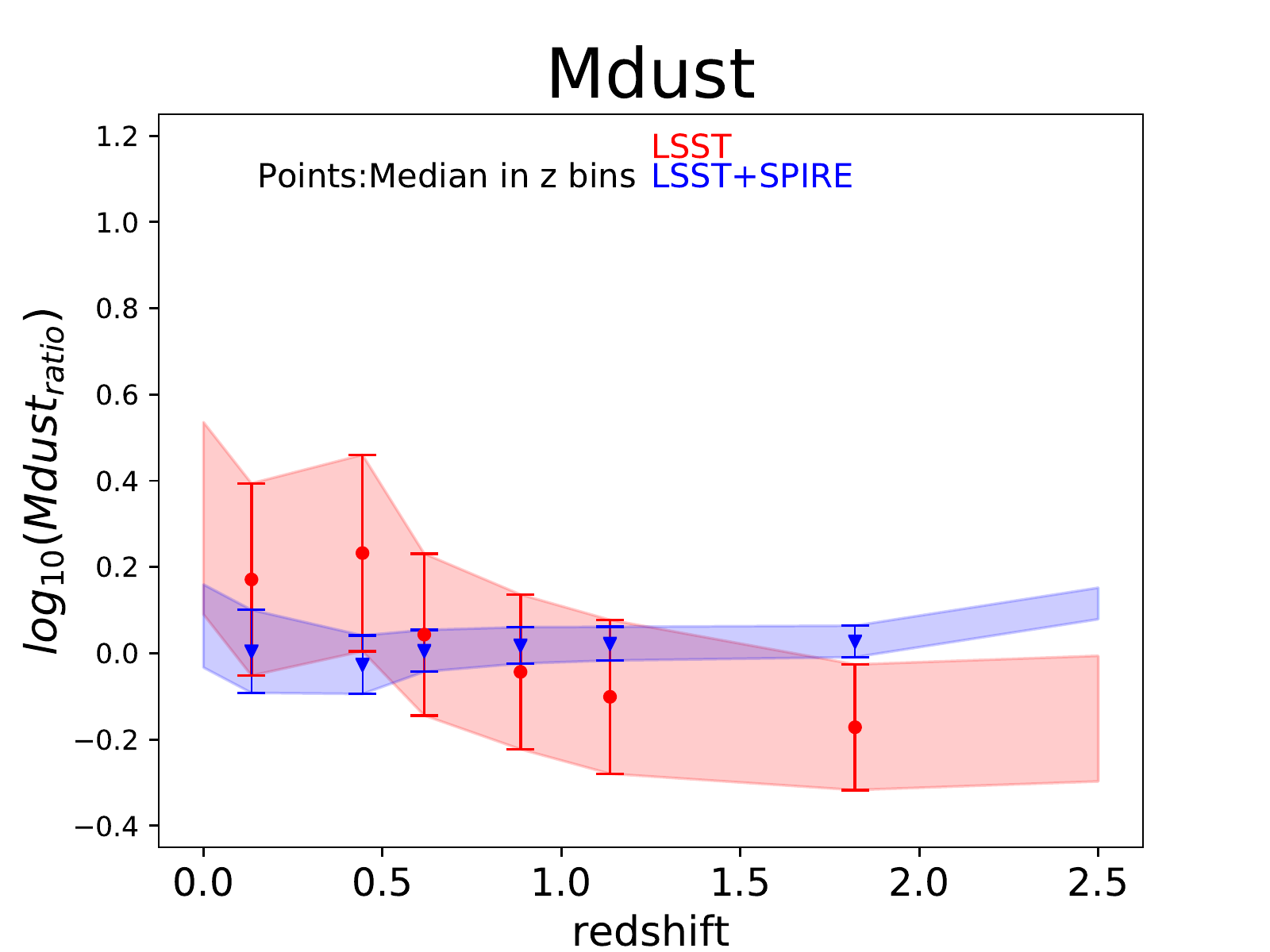}
    \caption{\textit{Upper left panel}: SFR ratio, defined as in figure \ref{fig:Density}, estimated with LSST observations only (red) and with LSST+IRAC observations (blue). The two samples are not in agreement at low redshift, but the comparison improves at high redshift, highlighting the result's redshift dependence. With the addition of MIR observations, the LSST-only estimates are consistent with the UV-IR ones. \textit{Upper right panel}: Same comparison but adding FIR SPIRE observations. In this case we do not fully remove the overestimation, highlighting that the problem lies in the lack of rest-frame MIR data. \textit{Lower left panel}: Same as in the other panels, but adding of GALEX observations. We obtain a slight decrease in the overestimation, but not enough to completely remove it, obtaining also an underestimation in the central redshift bins. For this comparison we consider the $\sim$ 3\,000 objects that have counterpart in the GALEX database. \textit{Lower right panel}: \mdust ratio evaluated with LSST observations only (red) and with LSST+SPIRE observations (blue). The results are consistent with the 'real' one with a very low scatter.}
    \label{fig:SFRratio_IRAC}
\end{figure*}

A different approach to correct the overestimation consists in applying the SED procedure on the LSST data together with other available observations in different bands (e.g., MIR \Spitzer bands, FIR \Herschel Spire bands). 
Figure~\ref{fig:SFRratio_IRAC} shows the results for SFR ratios obtained by adding IRAC MIR and SPIRE FIR observations. 
We expect that by adding the rest-frame NIR part of the SED will better constrain the attenuation of the old stellar population, while the MID and FIR mainly constrains the dust emission from the star-forming regions and the missing SFR, hidden by dust. 
The left upper panel of Fig.~\ref{fig:SFRratio_IRAC} shows that, just adding MIR observations, the overestimation of the SFR is fully corrected independently of the considered redshift range. 

We could also expect that the combined use of UV and LSST data will correct the overestimation of the SFR at low redshift. 
We test this hypothesis by adding the UV observations from GALEX and performing the SED fitting.
We perform a 1.5 arcsec crossmatch with the HELP catalogue and we identify $\sim$3\,000 galaxies having a GALEX counterpart. 
Figure~\ref{fig:SFRratio_IRAC} (left bottom panel) shows the comparison  of the SFR estimation for LSST-like and UV--FIR data set of those 3\,000 galaxies.  
We conclude from this plot, that by adding GALEX observations we obtain in general slightly lower overestimation but still consistent with previous results, meaning that the observed UV fluxes are not enough to completely correct the differences. Furthermore for $0.5<z<1.5$ we obtain a slight underestimation of the parameter. 
Nevertheless, we  stress that this result can be biased due to the low number of GALEX counterparts of the LSST-like  catalogue. 
and the low quality of GALEX observations for higher redshift sources. To confirm this statement, we simulate GALEX NUV and FUV observations for the whole sample (from now we refer to it as $GALEX_{true}$), again using the CIGALE module \textit{fluxes} , and employ it togheter with LSST to estimate the physical properties. We decide to cut objects with $z>1.5$ as, after this limit, both NUV and FUV GALEX bands are probing emissions below the Lyman break. Fig. \ref{fig:galex_corr} shows the comparison of the SFR  evaluated in this way and the one evaluated using LSST observation only.
We find a clear correction of the overestimation, highlighting the great impact that UV observations have on the SFR estimation. We confirm that the overestimation is partially due to the lack of the direct tracer of the young stellar population. Unfortunately, as there are no new UV missions planned in the near future, we do not expect to have available a UV coverage large enough to be used in conjunction with LSST in order to correctly estimate the physical parameters.

Figure~\ref{fig:SFRratio_IRAC} (right bottom panel) also shows the correction of \mdust estimates obtained by adding the SPIRE FIR observations.  
The good agreement obtained between the LSST plus SPIRE and the real estimates is due to the fact that the FIR emission is a direct probe of the dust mass.

\section{Testing different input parameters in CIGALE}
\label{different_param}

As discussed in section~\ref{Subsec:CIGALE}, the set of parameters employed for the analysis presented so far corresponds to the best one to fit the large sample of objects in the area of  $\sim$1\,300 deg$^2$ of the HELP field.
However, we also investigate how much the results obtained by only fitting the LSST simulated data are dependent on the CIGALE input parameters.
In particular, we tested possible variations of the derived galaxy physical quantities as a function of the input radiation field, PAH fraction and dust attenuation law.
For testing how the variations in the dust attenuation laws change our results, we re-fitted the UV to IR photometry and we re-derived the LSST simulated data that were later re-fitted by adopting a different attenuation law.

\subsection{Dust emission and mass}

Dust continuum emission is only determined by the energy balance and therefore it only depends on the amount of absorbed radiation. As a consequence, the total dust emission is not affected by the parameter $U_{min}$, but it is only sensitive to the total absorbed radiation.
On the contrary, we find that \mdust is not constrained by using the LSST data only, and its estimate is largely impacted by $U_{min}$ parameter employed. Indeed, by changing $U_{min}$ the amount of radiation that irradiates the dust is modified, but the amount of dust emission is unaltered. As a consequence, larger input values of $U_{min}$ are translated in lower \mdust and vice versa.
Large values of the $U_{min}$ ($U_{min}$=25) parameter yield an underestimation of \mdust.
As a consequence, since the shape of dust emission is completely unconstrained using only the LSST coverage, the Bayesian method cannot evaluate the $U_{min}$ parameter and assigns to all galaxies an average value among all the input parameters considered. This results in having dust mass estimates strongly dependent on the input parameters employed. When the dust emission is constrained using SPIRE observations, the $U_{min}$ parameter is well evaluated by the Bayesian method, and so the dust mass.

The use of different attenuation laws also changes the estimates of dust emission and mass, since the radiation absorbed and the re-emitted by dust grains is modified. 
By using either the \citealt{Calzetti2000} and \citealt{charlot2000} attenuation laws, we obtain an overestimation of the dust luminosity over the entire redshift range. The trends between the dust luminosity ratios and the redshift are however different.
As far as the dust mass is concerned, by adopting \citealt{Calzetti2000} we obtain a constant slight overestimation of the dust mass, centred around 0.3 dex across the whole redshift range, while by using \citealt{charlot2000} the mass of dust is overestimated for local galaxies and underestimated for redshift greater than $\sim 1$. 

\subsection{Dust attenuation laws and the sfar formation history}
\begin{figure}
    \centering
    \includegraphics[width=0.9\hsize]{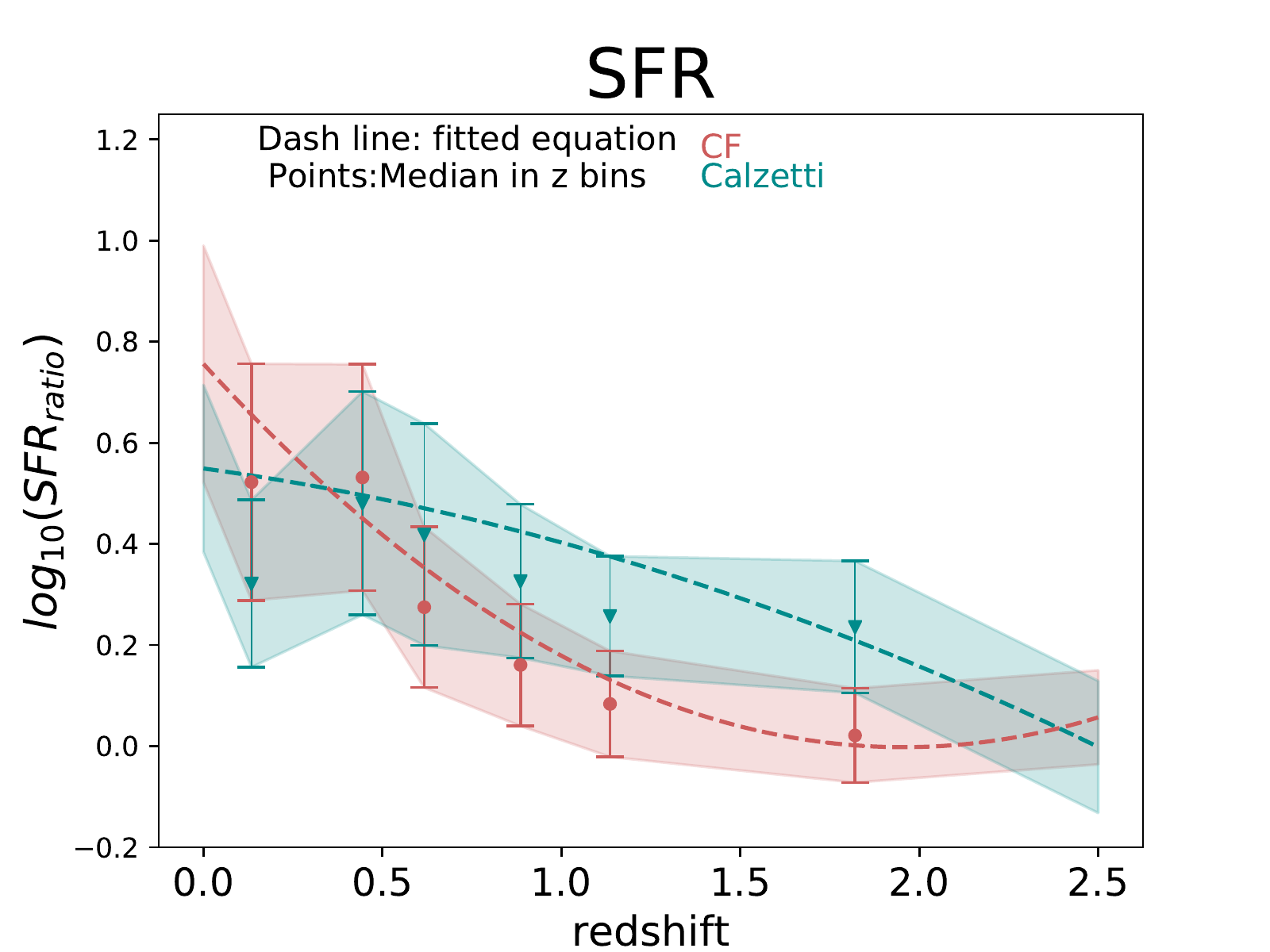}
    \caption{The SFR overestimation in function of redshift for  \citealt{charlot2000} (red) and \citealt{Calzetti2000} (cyan) attenuation laws. }
    \label{fig:calzetti}
\end{figure}

We find the SFR to be unaffected by the PAH fraction and by the input radiation field ($U_{min}$ and $\alpha$ parameters) since the aforementioned input quantities only shape dust emission.
The SFR is instead influenced by our choice of the attenuation law. When the \cite{Calzetti2000} modified attenuation law is employed, we obtain the results shown in Fig.~\ref{fig:calzetti}. The results using \cite{Calzetti2000} attenuation curve are in good agreement with the ones obtained with the \cite{charlot2000} prescription at low redshift but change in shape going to higher \textit{z}. The overestimation with the  \cite{Calzetti2000} curve is more constant along the redshift range, while  it decreases faster when using the \citealt{charlot2000} prescription.

Furthermore, we check also the dependence of SFR estimates on the SFH module used. Performing the entire process using a delayed plus additional burst SFH we obtain an even higher overestimation when estimated using just LSST observations, that again decreases going to higher redshift.


We also inspect changes in the SFR difference while using the best values, instead of Bayesian one, for the LSST-like sample. We find a consistency between the SFR from the best fit and the one estimated from the UV-FIR fit, finding $log_{10}(SFR_{LSST, best}/SFR_{UV-FIR})$ very well distributed around zero with a scatter of 0.05 dex. This result, that shows a good agreement between the best templates of the two runs, is understandable as the fluxes used for the LSST-like run were calculated based on the best template of the UV-FIR run.
We confirm that the differences found must be sought in the Bayesian analysis, that tends to overestimate the attenuation when employed on LSST data only. Unfortunately, being estimated directly from the SED that best fits our data, the best-fit value has several drowbacks that make it not suitable for this type of analysis. For example it ignores the degeneracies one can encounter, as models with equally good fits can have very different properties. Moreover, the best fit in itself does not provide information on the uncertainties. 

\section{Application of the A$_{FUV}$-\mstar relation to correct the SFR overestimation }
\label{Sec:AFUVMSTAR}

Previous results shows major miscalculations of the SFR while using the LSST data only. Due to the lack of information in the UV and MIR part of the spectrum, the SED fitting results in an overestimation of the attenuation, that leads to the general overestimation of the SFR. 
At the same time, as it is probed mainly from the optical emission of the galaxy, the \mstar seems to be well estimated using only LSST data. 
Taking into account that the \mstar is the result of the previous star formation activity of the galaxy, which is responsible for producing the dust, it could be used as a promising tracer of the dust content. 

During past years, several works have explored possible relations between \mstar and dust attenuation (\citealt{Xu2007}, \citealt{Martin2007}, \citealt{Buat2009}, \citealt{Jana2020}). 
Most of them suggest 
a possible linear relation between \AFUV and $log_{10}(M_{star})$ over a large mass range ($9 \leq log_{10}(M_{star}) \leq 12$). 
According to the literature this relation is highly dependent on redshift.

Recently, \cite{Jana2020}, hereafter \citetalias{Jana2020}, modelled a single parameter linear function, assuming a non-zero constant dust attenuation for low mass galaxies (Eq. 6, in \citetalias{Jana2020}). 
They used a sample of galaxies based on the selection criteria which requires the IR Excess\footnote{The IR Excess is defined as IRX=log($L_{IR}/L_{UV}$), where the $L_{IR}$ stands for total integrated luminosity in the IR, and the $L_{UV}$ is the UV luminosity  derived from flux measured with a filter,
such as, for e.g. GALEX or estimated via SED process.} (IRX)  calculated directly from the  IR-to-UV ratio or by SED fitting. 
This selection can introduce a bias in the local Universe and above redshift 2--3, due to  the IR detection (at high redshift only the very dusty and massive galaxies are detected, while in the local universe the IR detected galaxies are rather rare). 
\citetalias{Jana2020} found that the \AFUV -- \mstar relation cannot be described with a simple linear function, and they conclude their work with a new relation between \AFUV and  \mstar in function of redshift. 
In this section we will try to use the \AFUV--\mstar relation  provided by \citetalias{Jana2020} to estimate the \AFUV  (from now on we will refer to it as  \AFUV$_{BB20}$) of the LSST sample.  
The procedure is as follows: (1) from the LSST data we estimate the \mstar, (2) using Eqs.~5 and 6 from \citetalias{Jana2020} we calculate the \AFUV$_{BB20}$, and finally (3) we use \AFUV$_{BB20}$ as a prior of the new LSST CIGALE run.

\begin{figure}
    \centering
    \includegraphics[width=0.9\hsize]{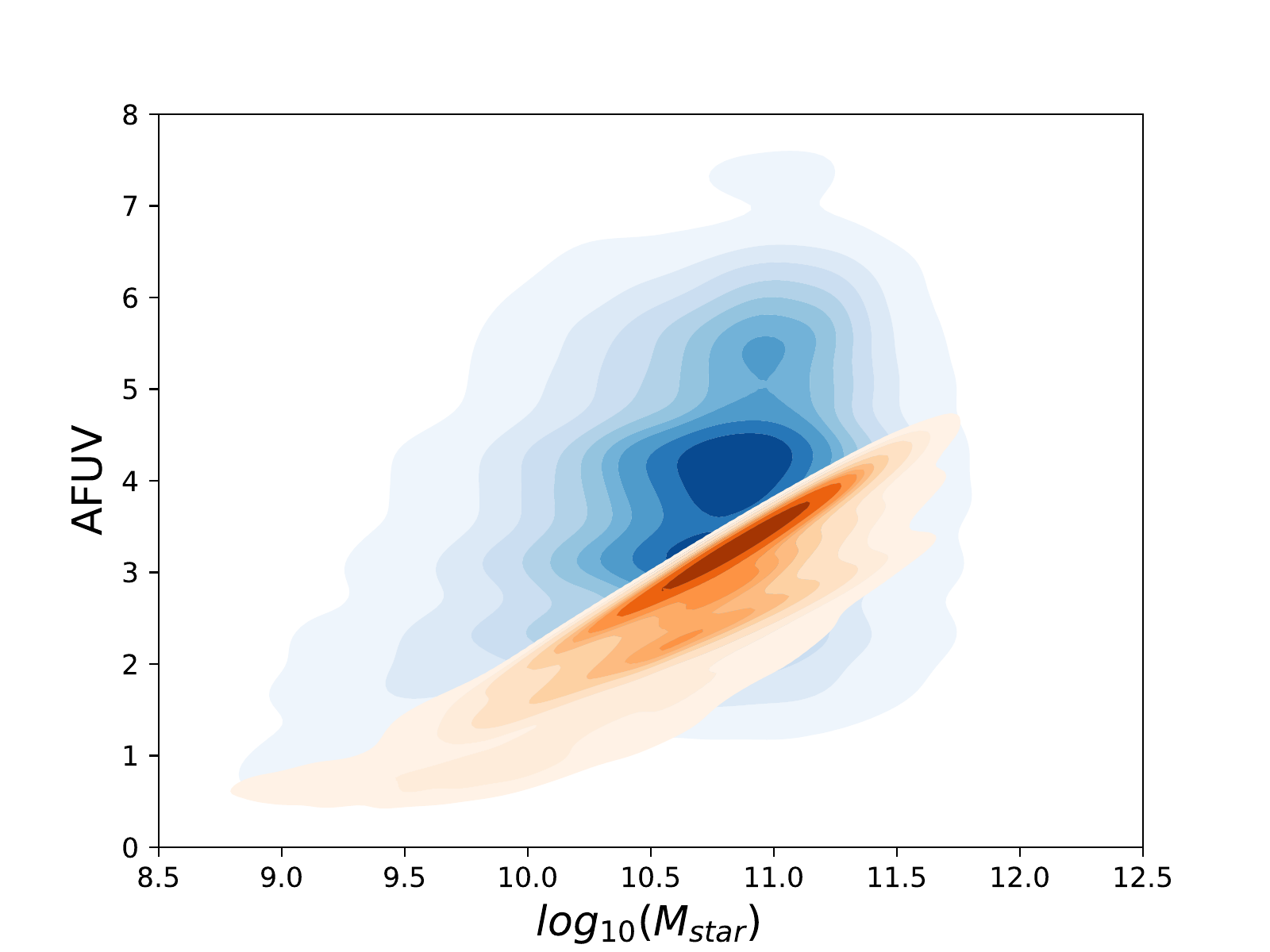}
    \caption{\AFUV in function of stellar mass estimated from the UV-FIR SED fitting (blue density plot) in comparison with the \AFUV estimated employing the $M_{star}$ in the relation from \cite{Jana2020}.}
    \label{fig:afuv_comparison}
\end{figure}

\begin{figure}
    \centering
    \includegraphics[width=0.9\hsize]{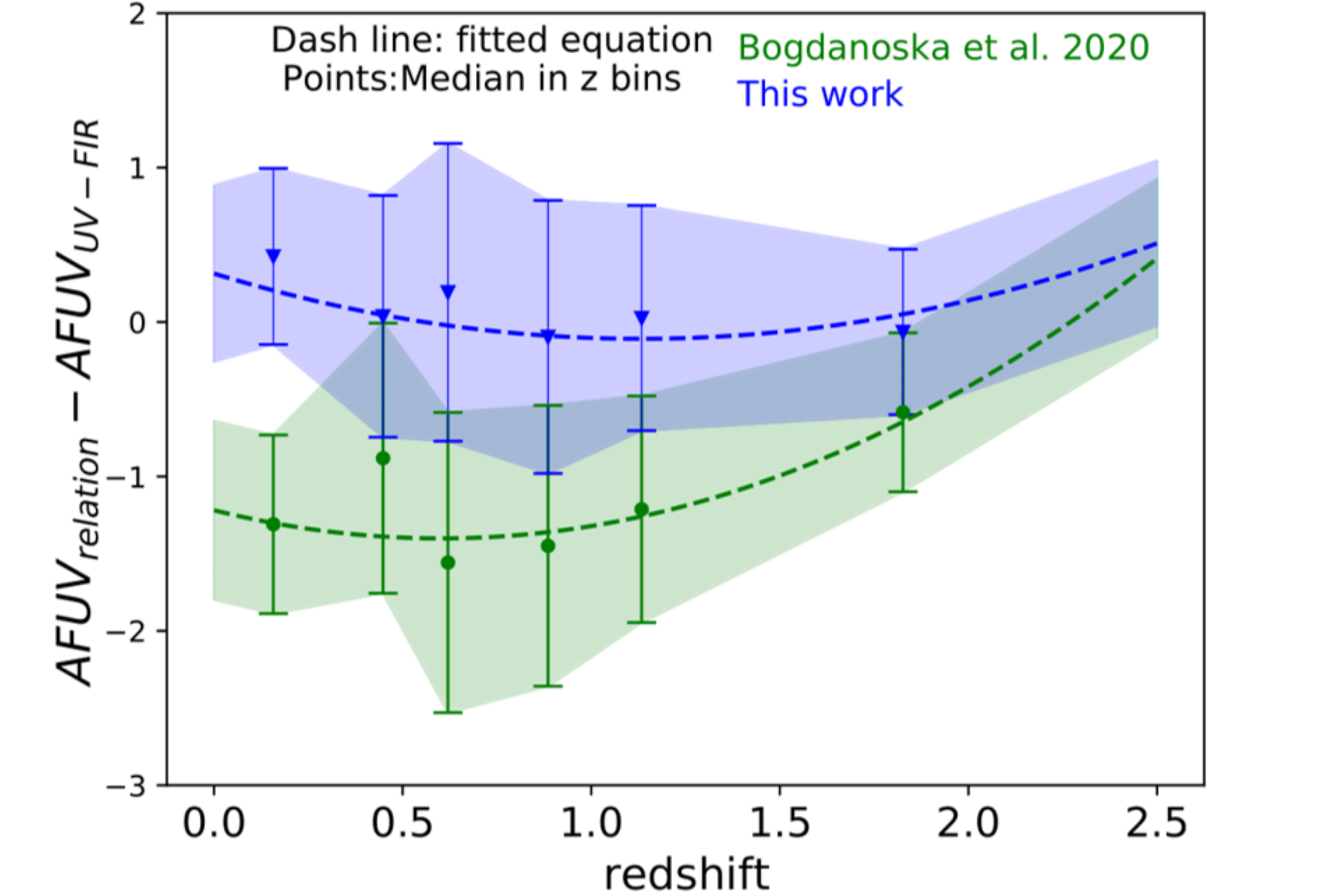}
    \caption{The difference between \AFUV estimated from \AFUV--\mstar relations:  \citealt{Jana2020} (green) and Eq.~\ref{eq:linear}  with coefficients reported in Table~\ref{tab:linear_fit} (blue), and \AFUV estimated from the UV-FIR SED fitting, in function of redshift.}
    \label{fig:att_difference}
\end{figure}

\begin{figure}
    \centering
    \includegraphics[width=1\hsize]{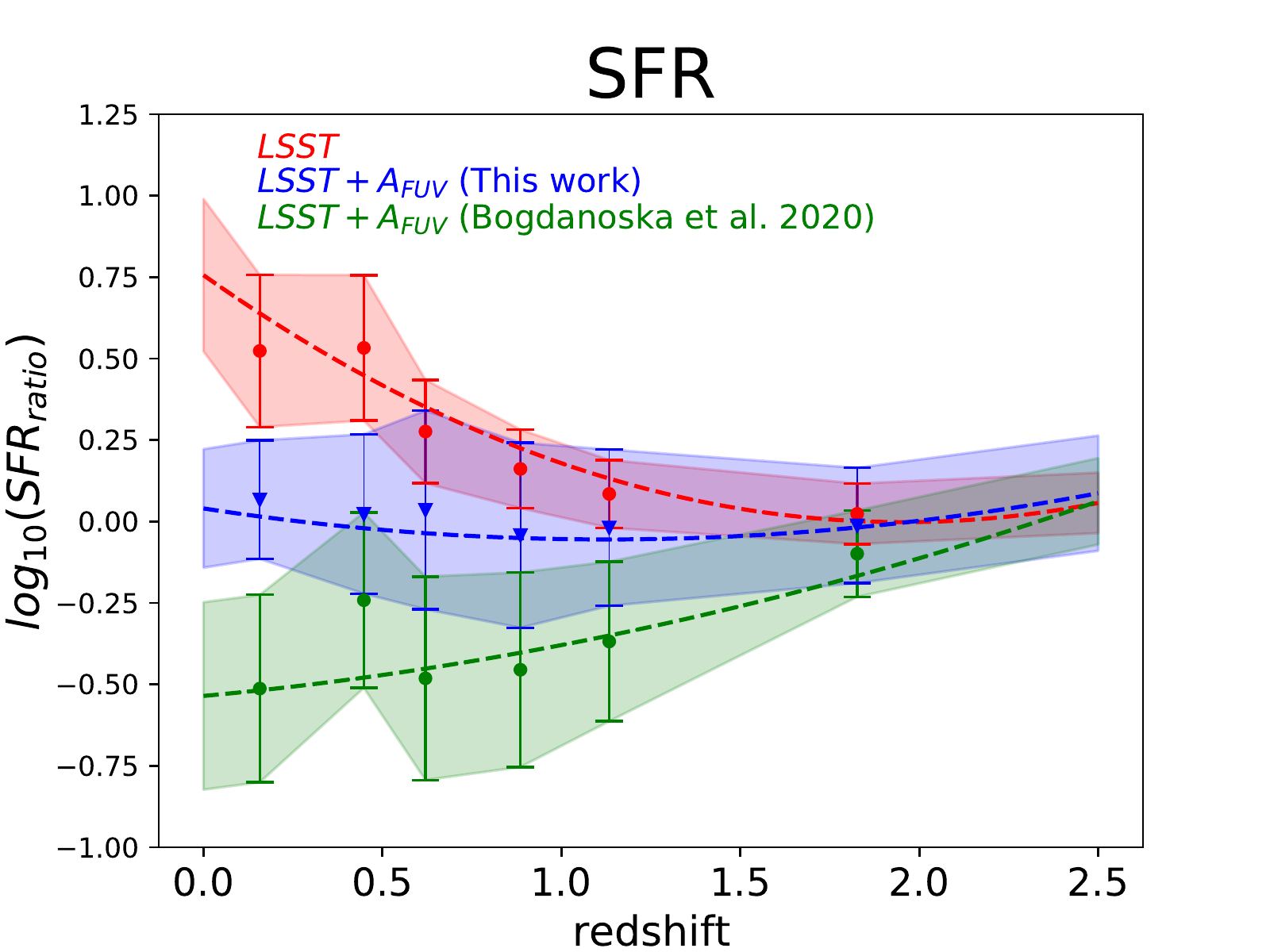}
    \caption{The SFR ratio, defined as in Fig.~\ref{fig:Density}, in function of redshift. 
    The result coming from the SED fitting employing only LSST data is shown in red. 
    Green and blue lines represent results from the SED fitting, adding as prior the \AFUV from both relations: \cite{Jana2020} and Eq.~\ref{eq:linear} respectively. }
    \label{fig:SFR_jana}
\end{figure}

Fig. \ref{fig:afuv_comparison} shows the \AFUV$_{BB20}$ 
in comparison to the \AFUV  obtained from the full UV-FIR SED fitting, in function of \mstar. We find that the estimates \AFUV$_{BB20}$ are substantially lower than those obtained with the SED fitting process. The difference between \AFUV$_{BB20}$ and \AFUV$_{UV-FIR}$ is shown in Fig.~\ref{fig:att_difference} (green line). 
We can notice that the underestimation is accentuated for low redshift objects. As expected, employing the \AFUV$_{BB20}$ as prior in the SED fitting process, along with LSST observations, result in an underestimation of the SFR, as shown in Fig. \ref{fig:SFR_jana}, where the ratios between the true SFR and the one derived by different methods are shown. 
We suspect that the reason for the substantial difference between the two estimates of \AFUV  can be traced back to the choice of the sample, as the sample used in \citetalias{Jana2020} is more general, while in our case we focus on IR bright galaxies, to ensure the highest quality of the UV-FIR SED fitting process.
It is clear from this results that we cannot employ directly \cite{Jana2020} relation to correct the SFR overestimation for our sample. 

We decided to incorporate the general idea presented in \cite{Jana2020} and   
to make use of a \AFUV--\mstar relation in order to correct the SFR for an LSST sample of data. 
We  build, albeit simplified, a relation that represents our sample of IR bright main sequence galaxies by following a procedure similar to \cite{Jana2020}. For this purpose, we fit the \AFUV estimated from the UV-FIR SED fitting as a function of the $log_{10}(M_{star})$.
We used four redshift bins (0-0.5, 0.5-1.0, 1.0-1.5, 1.5-2.5) to include the redshift dependence of the \AFUV--\mstar relation.
Linear, power law and exponential functions were tested to obtain the best fit, but we found a negligible difference between them. 
To be as consistent as possible with the results obtained in the previous works, we decided to use the linear function, in the form: 
\begin{equation}
\label{eq:linear}
    A_{FUV-LSST}=a \cdot log_{10}(M_{star-LSST})+b.
\end{equation}
From the fitting process, we obtain  $a$ and $b$ coefficients, for each redshift bin. 
Table~\ref{tab:linear_fit} shows all coefficients together with uncertainties.

\begin{table}
\caption{Obtained $a$ and $b$ coefficients from  fitting Eq.~\ref{eq:linear}, in the four redshift bins. }    
\label{tab:linear_fit}      
\centering                          
\begin{tabular}{c c c}        
\hline\hline                                       
Redshift & a & b   \\\\ 
\hline
0-0.5 & $0.41 \pm 0.02$ & $-1.39 \pm 0.21$ \\
0.5-1 & $0.44 \pm 0.03$ & $-0.49 \pm 0.30$ \\
1-1.5 & $0.72 \pm 0.03$ & $-3.42 \pm 0.34$ \\
1.5-2.5 & $0.83 \pm 0.04$ & $-5.19 \pm 0.39$ \\
\hline
\end{tabular}
\vspace{1mm}
\end{table}

Blue line in Fig.~\ref{fig:att_difference} shows the difference between the $A_{FUV-LSST}$ calculated with our four linear relations and the one estimated from the UV-FIR SED fitting, as a function of redshift. 
We can see that our relation better reproduce the AFUV derived from the fitting of the full SED than  \cite{Jana2020} relation. By employing the $A_{FUV-LSST}$ as a prior in the SED fitting, along with LSST observations, we obtain the blue relation shown in Fig. \ref{fig:SFR_jana}. 
This figure shows that the SFR overestimation is fully corrected while applying $A_{FUV-LSST}$ prior.
This result also proves that, knowing the \AFUV--\mstar relation for a given sample of galaxies, it is possible to estimate SFR without the IR counterpart.    
This requires prior knowledge of the sample, which is often not available. 
We are aware that the relation constructed in this work may be applicable only to our sample, or at most to IR-bright normal, star forming galaxies, but further generalization of 
our results is outside the scope of this paper. 
However, considering the extreme usefulness of this relationship for future surveys as LSST,  we are planning to extend our analysis to a more general sample of galaxies in the next work.

\section{Conclusions}
\label{Sec:Conclusions}

We perform a reliability check of physical properties estimation of MS galaxies by employing simulated LSST observations. 
For this purpose we select respectively 50\,135 and 15\,754 objects from \EN and COSMOS fields of the \textit{Herschel} Extragalactic Legacy Project  (HELP), in order to build the starting set of data to simulate observed LSST fluxes and to obtain reliable estimates of the physical properties of galaxies. 

An important part of our analysis is the sample selection. We selected only galaxies from so called main sequence by removing all possible SBs from the sample using the same method as  \cite{Rodighiero2011}, and passive galaxies using the method described in \cite{Salim2018}. Furthermore we removed also galaxies that contain AGN according to \cite{Stern2005} selection.
We also cleaned the sample from all non-typical galaxies by  implementing  additional quality criteria on physical properties following \cite{Malek2018}: from our analysis we removed all galaxies for which the \ldust and \mstar estimated from the full SED fitting (from UV to FIR) are different from the ones obtained from only the optical or only the infrared part of the spectrum. 
At the end of the sample selection, we selected in total 43\,652 galaxies, 86\% of the total sample. 

 We used such a  sample of MS galaxies as a prior for the calculation of the corresponding LSST fluxes in the \textit{ugrizy} bands. 
 We used the LSST simulation software package \textit{CatSim}, in order to simulate the uncertainties on the photometric measurements, we took into account the possible effects due to the hardware and observational components (e.g. detector, dark sky, atmosphere). 
 We then estimated the main physical properties of galaxies by performing the SED fitting of the simulated LSST data by employing the same sets of modules and parameters as for the originally used HELP galaxies (Shirley et al. in prep., \citealt{Shirley2019}).

  We found that \mstar is well estimated by the LSST--like data set. 
  At the same time SFR, \ldust are overestimated using the LSST--like sample only, while \mdust is completely unconstrained and dependent on the input parameters employed.
  The overestimation of the SFR is redshift dependent, and clearly decreases with redshift, disappearing around redshift~$\sim$1. 
  We found the relation which can correct the overestimation for the SFR parameter: 
  $\rm \log_{10}(SFR_{ratio})=0.26\cdot \textit{z}^2 -0.94\cdot \textit{z} +0.87$.

  We check which photometric data can be combined with the LSST data to remove the overestimation.    
  In our analysis we used not simulated but real, sometimes uncompleted, data to fully mimic the auxiliary data for the LSST one, since we do not expect to have better UV  or FIR data soon. 
  We found that the most efficient way to correct the overestimation of the SFR is by adding mid-IR observations (IRAC data), while \mdust is corrected by adding the far-IR bands (SPIRE data). The addition of UV observations from GALEX does not correct the differences. 
  Our findings suggest that the main problem of the pure LSST-like sample in the Local Universe will be the inability to mimic the real attenuation for the old and young stellar populations.

  By testing the input parameters of CIGALE, we found that the SFR overestimation is preserved using different attenuation laws commonly employed in the literature (e.g. \citealt{Calzetti2000}, \citealt{charlot2000}), but its trend as a function of the redshift changes. The estimate of \mdust is instead found to be dependent on both the input radiation field ($U_{min}$) and attenuation law and is unconstrained if LSST data only are employed for the SED fitting.
  
  In Section~\ref{Sec:AFUVMSTAR} we show that another efficient  way to  correct the SFR is by exploiting, if available, a prior knowledge of \AFUV. 
  We stress that the further  analysis of the \AFUV--\mstar relation can be useful for a future surveys and help to properly estimate main physical parameters of galaxies without IR observations. 
  
  As future work, we plan to extend this test to different SED fitting methods and HELP fields, to check the systematics of our  results.
  
 \begin{acknowledgements}
 We would like to thank William Pearson for his help in understanding our object's behaviour in the SFR-\mstar diagram. 
 GR, KM, AN, MH and AP acknowledges support from the National Science Centre (UMO-2018/30/E/ST9/00082, UMO-2018/30/M/ST9/00757 and UMO-2020/38/E/ST9/00077). 
Authors are grateful for the support from Polish Ministry of Science and Higher Education through a grant DIR/WK/2018/12. We also
thank the anonymous referee who has helped clarify and improve various aspects
of this article.
 \end{acknowledgements}

\bibliographystyle{aa} 
\bibliography{bibliography.bib}

\begin{appendix}

\section{Outlier selection}
\label{app:outlier}
\begin{figure}[htb]
    \centering
    \begin{tikzpicture}
        \node[anchor=south west,inner sep=0] (image) at (0,0) {\includegraphics[width=1\hsize]{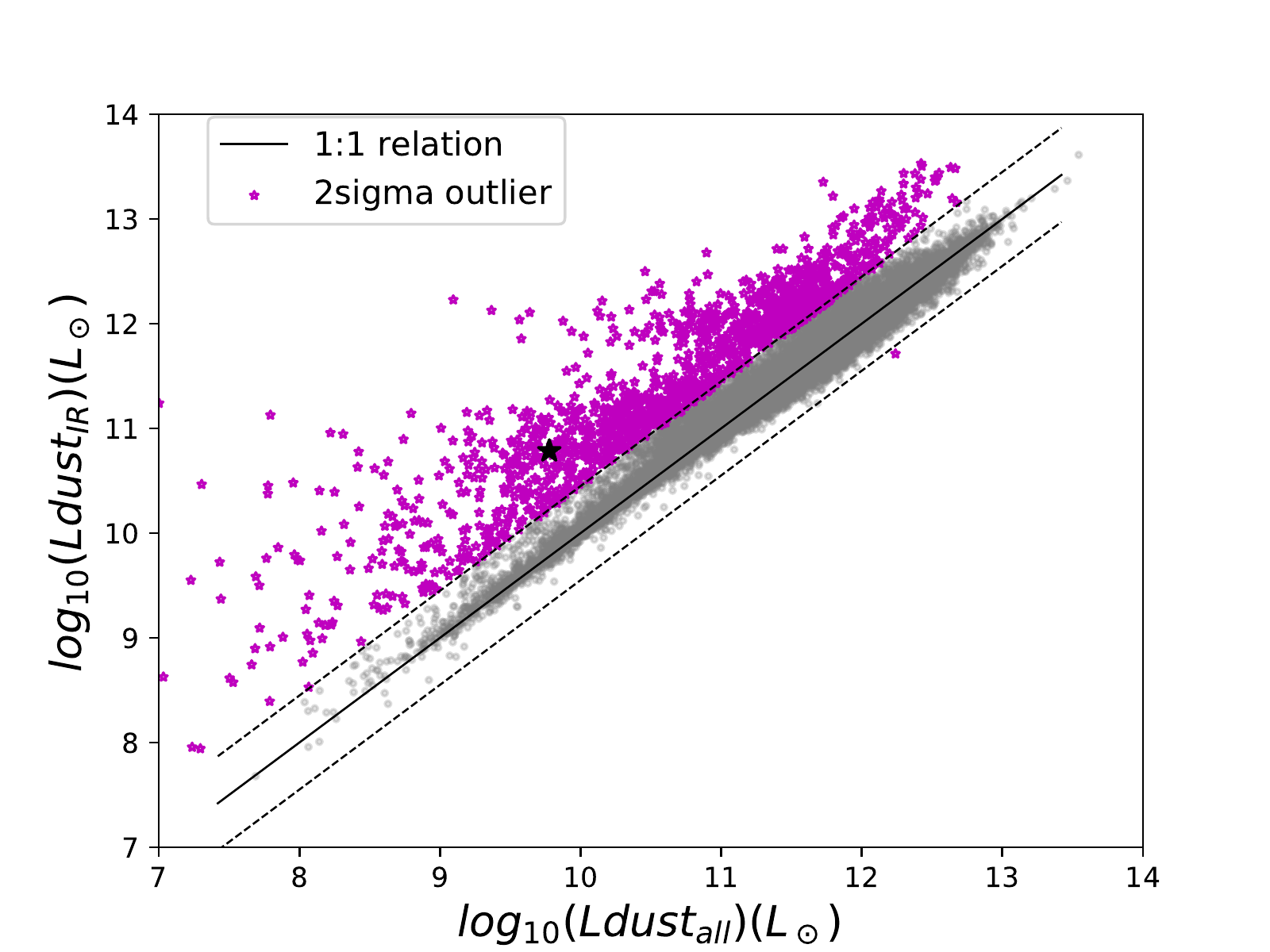}};
        \begin{scope}[x={(image.south east)},y={(image.north west)}]
            \node[anchor=south west,inner sep=0] (image) at (0.52,0.11) {\includegraphics[width=0.38\hsize]{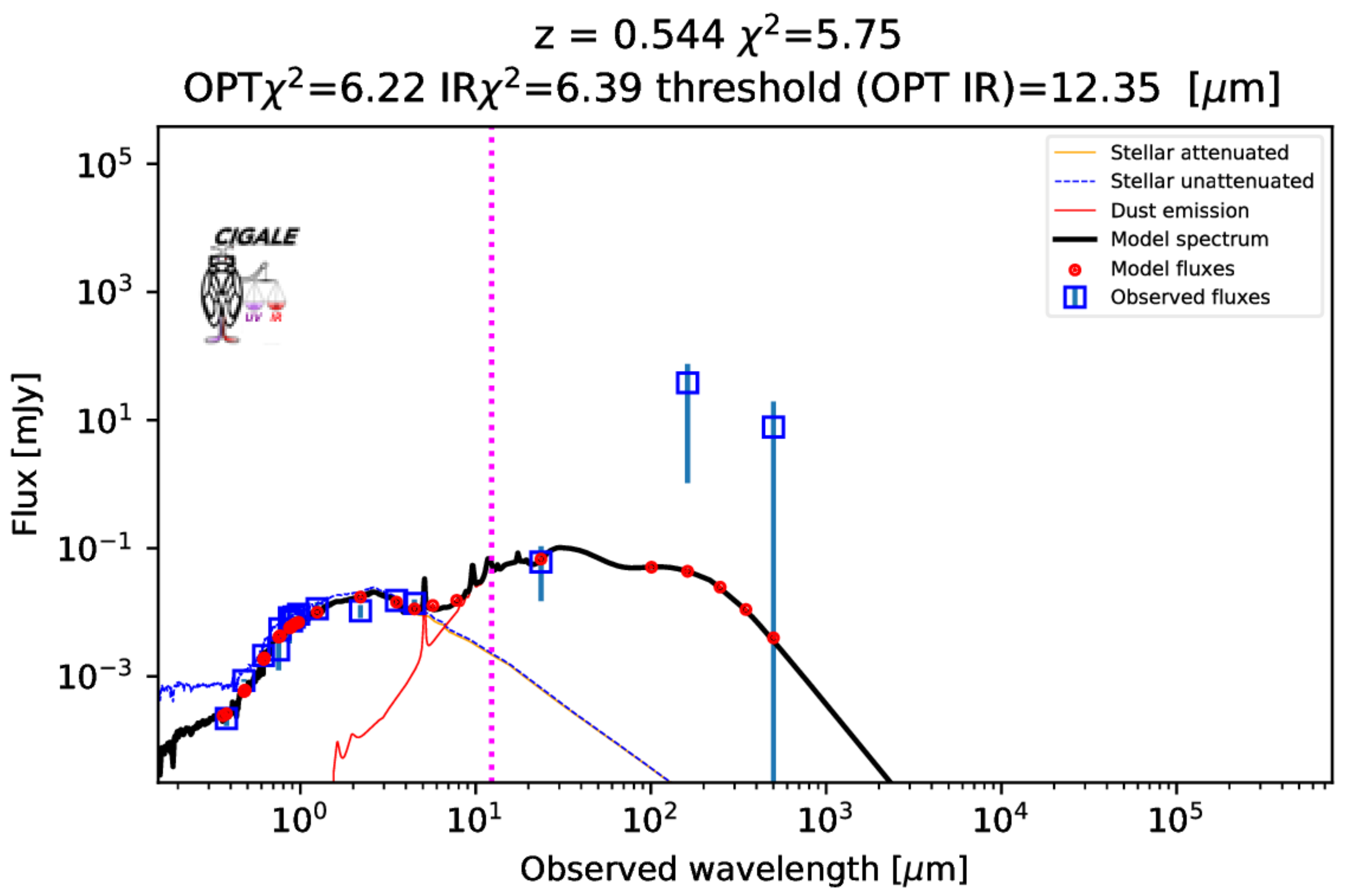}};
        \end{scope}
    \end{tikzpicture}
    \begin{tikzpicture}
        \node[anchor=south west,inner sep=0] (image) at (0,0)
        {\includegraphics[width=1\hsize]{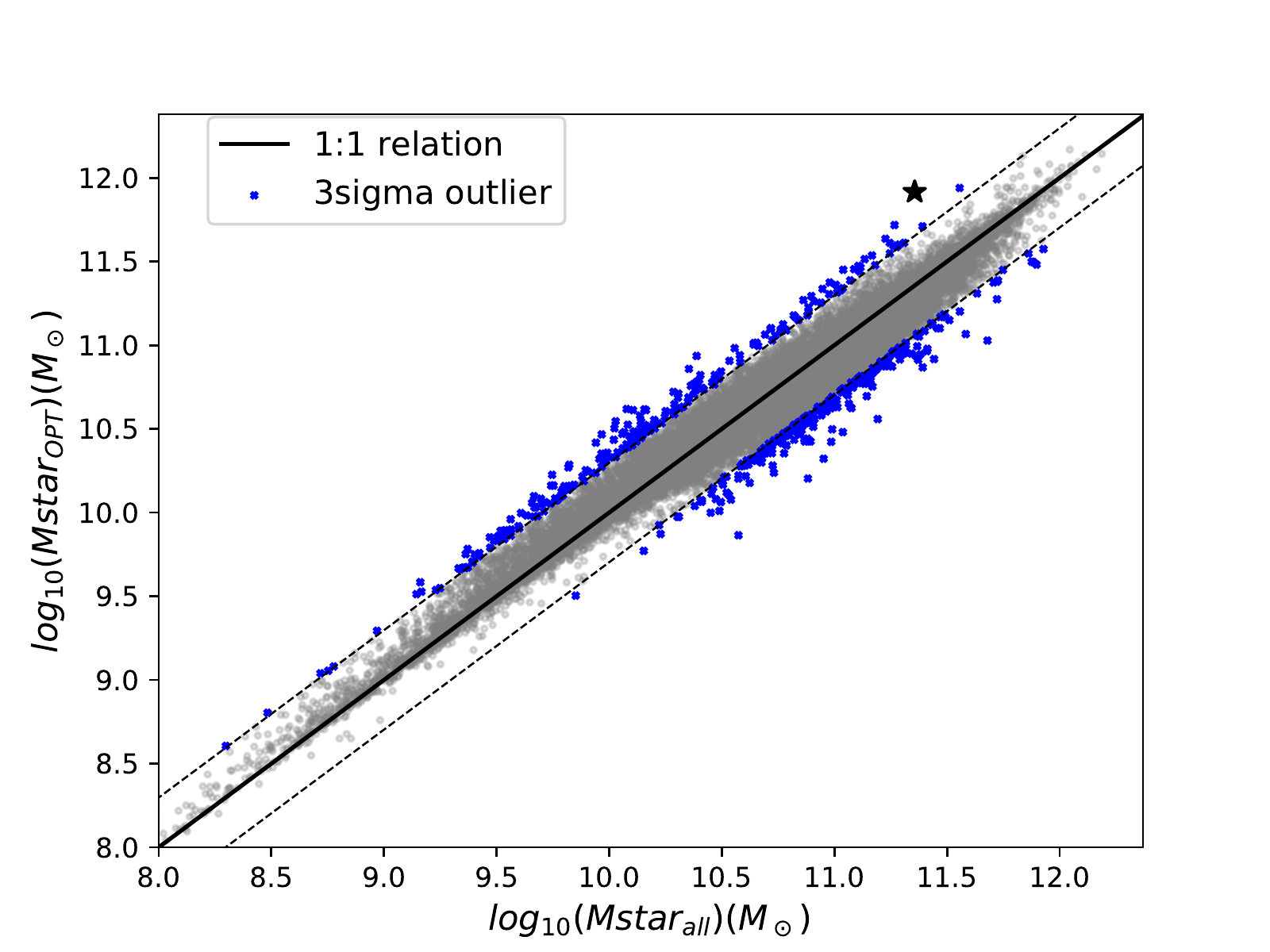}};
        \begin{scope}[x={(image.south east)},y={(image.north west)}]
            \node[anchor=south west,inner sep=0] (image) at (0.54,0.12) {\includegraphics[width=0.35\hsize]{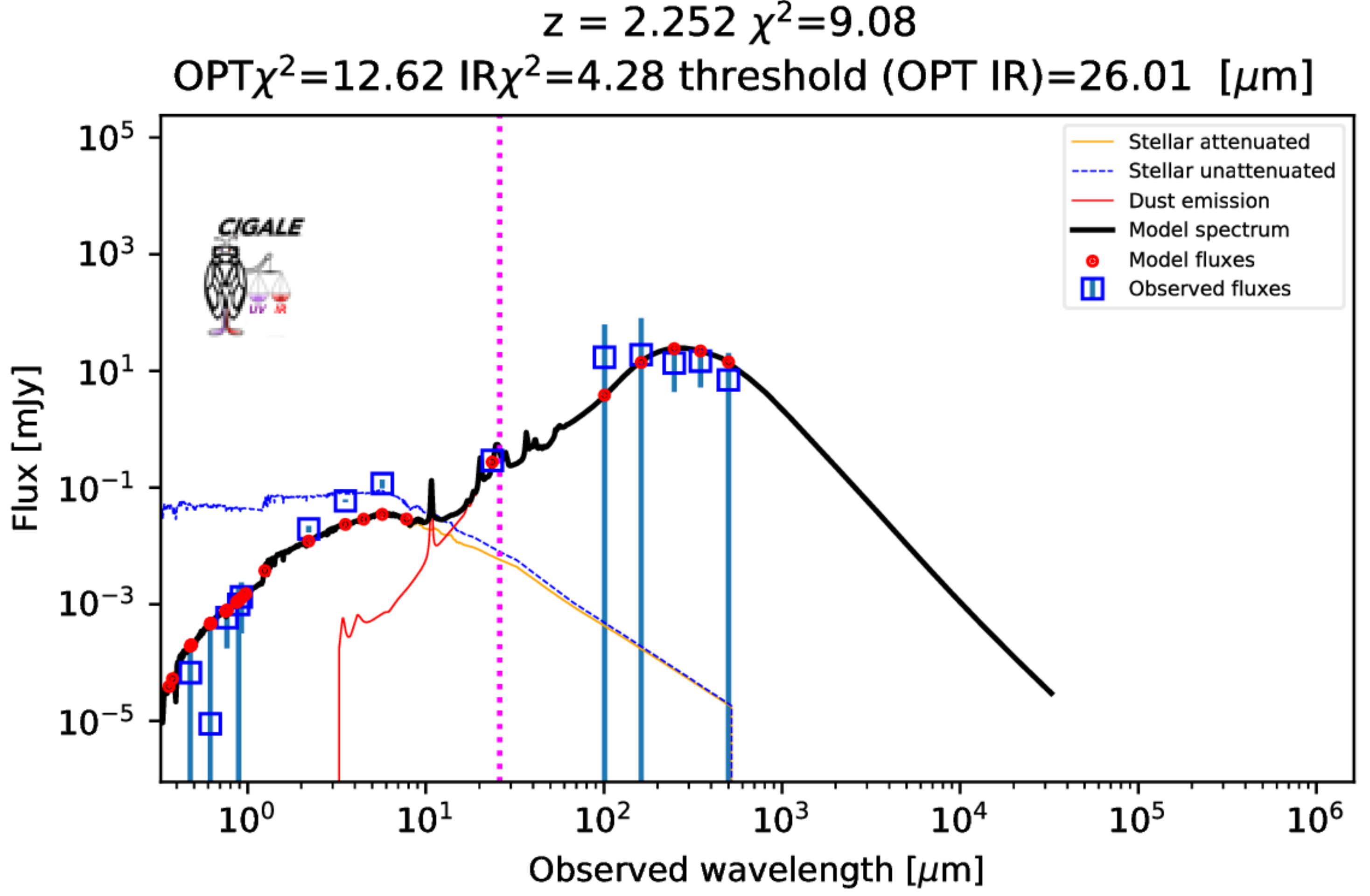}};
        \end{scope}
    \end{tikzpicture}
    \caption{Outlier selection. Comparison between  $L_{dust,IR}$ and  $L_{dust,all}$ (\textit{upper panel}), and  $M_{star,OPT}$ and $M_{star,all}$ (\textit{bottom panel}). The \ldust inconsistent objects are represented as magenta stars, while full blue stars correspond to the \mstar outliers. Gray circles represent objects with consistent estimate of the \ldust and \mstar  parameters. Inside each panel we present an example SED of the respective black star highlighted outliers. }
    \label{fig:out_selection}
\end{figure}

To implement additional quality criteria based on physical properties of the sample, we run CIGALE two more times: (1) for optical data only, to estimate the stellar mass based  on the optical measurements only (from now $M_{star,OPT}$), and (2) for FIR data only, to calculate the dust luminosity (hereafter $L_{dust,IR}$). 
In Fig.~\ref{fig:out_selection} we compare the physical properties obtained by employing this method with the one from full wavelength, UV-FIR, fits ($M_{star,all}$, and $L_{dust,all}$, respectively). 

Based on that analysis we eliminate galaxies that show \ldust and/or \mstar inconsistency with the ones estimated from the full SED fitting.
Outliers are selected based on the distance from the 1:1 relation:

\begin{itemize}
    \item criterion 1: $L_{dust}$ inconsistent (within $2\sigma$ level) with the $L_{dust,IR}$
    \item criterion 2: $M_{star}$ inconsistent (within $3\sigma$ level) with the $M_{star,OPT}$
\end{itemize}

As shown in \cite{Malek2018}, the inconsistency between estimated \ldust values might be  induced by energy balance issues of heavily dust-obscured galaxies or lensed objects. 
We have also removed galaxies with inconsistent \mstar estimation, mostly due to problems with optical and IR catalogues matching. 
Figure~\ref{fig:out_selection}  shows also two example SEDs for objects considered as outliers. Based on  these  criteria we remove 1\,642 \EN sources (4.5\% of the total sample), and 460 COSMOS sources (3.2\% of the total sample) with inconsistent estimates of \ldust. 
The \mstar inconsistency removed  475 (1.3\% of the total sample), and 214 (1.5\% of the total sample)  galaxies for \EN and the COSMOS field respectively. 

\section{AGN selection}
\label{app:AGN}

\begin{figure*}
    \centering
    \includegraphics[width=0.4\hsize]{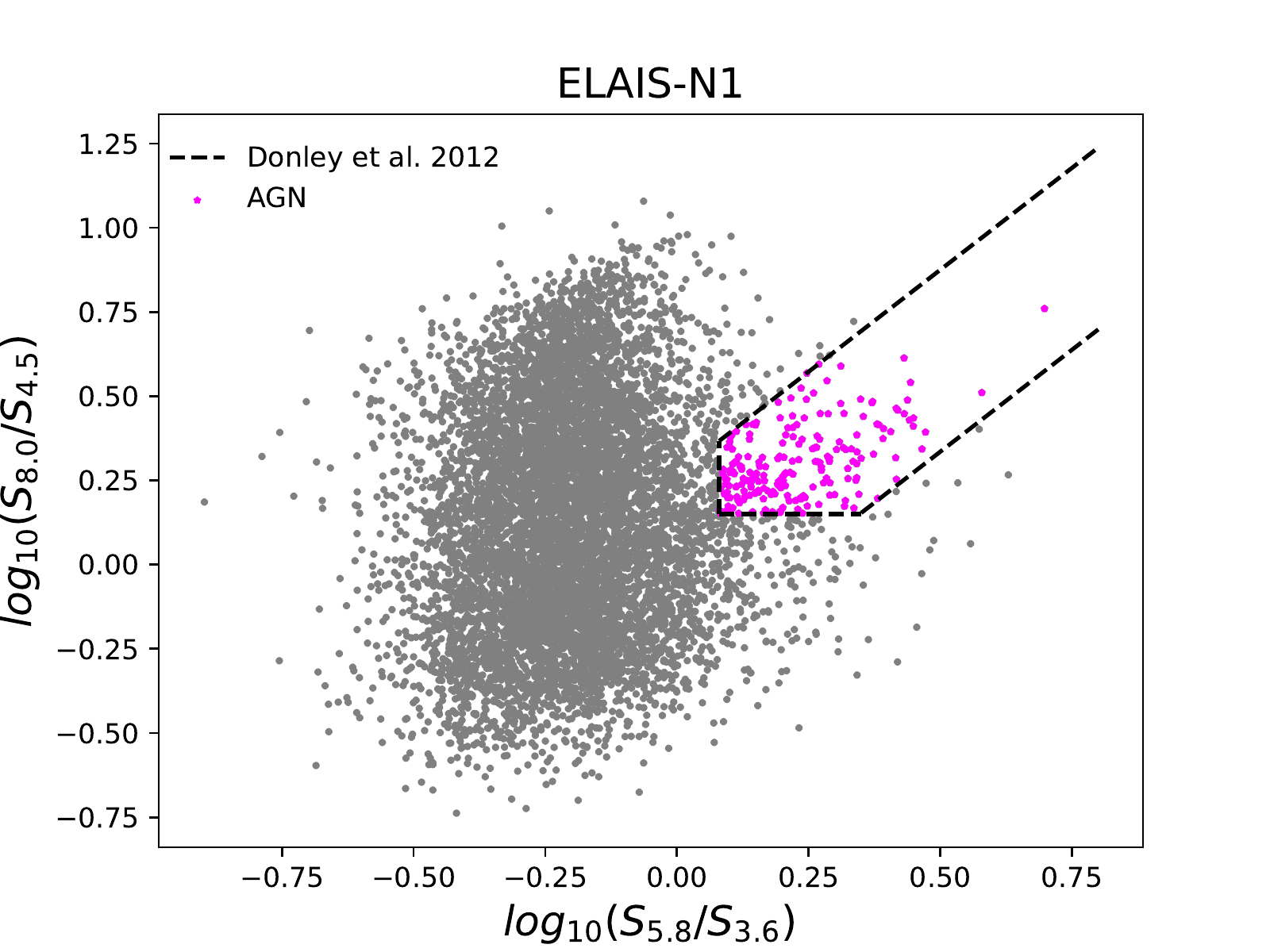}%
    \includegraphics[width=0.4\hsize]{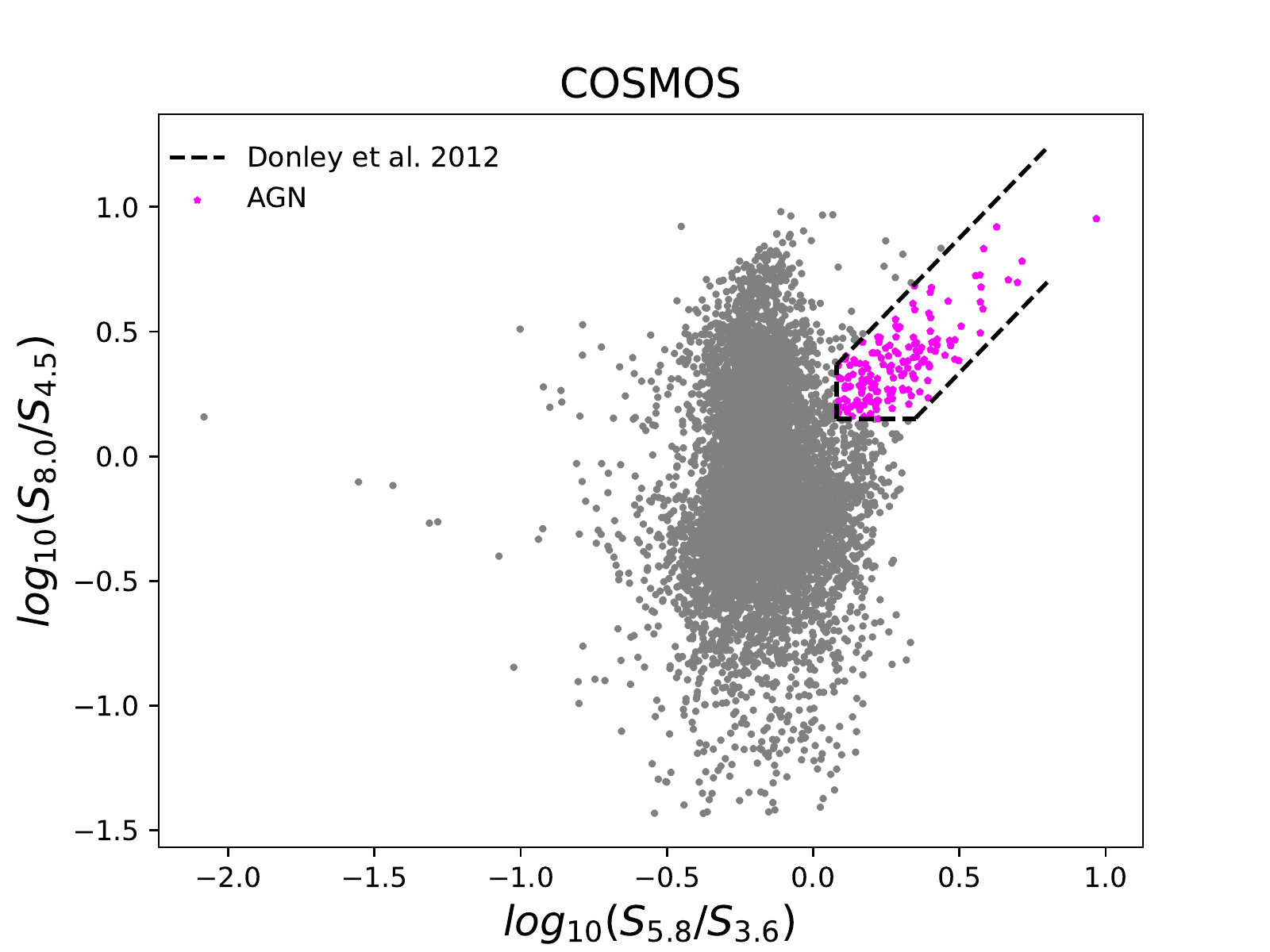}
    \includegraphics[width=0.4\hsize]{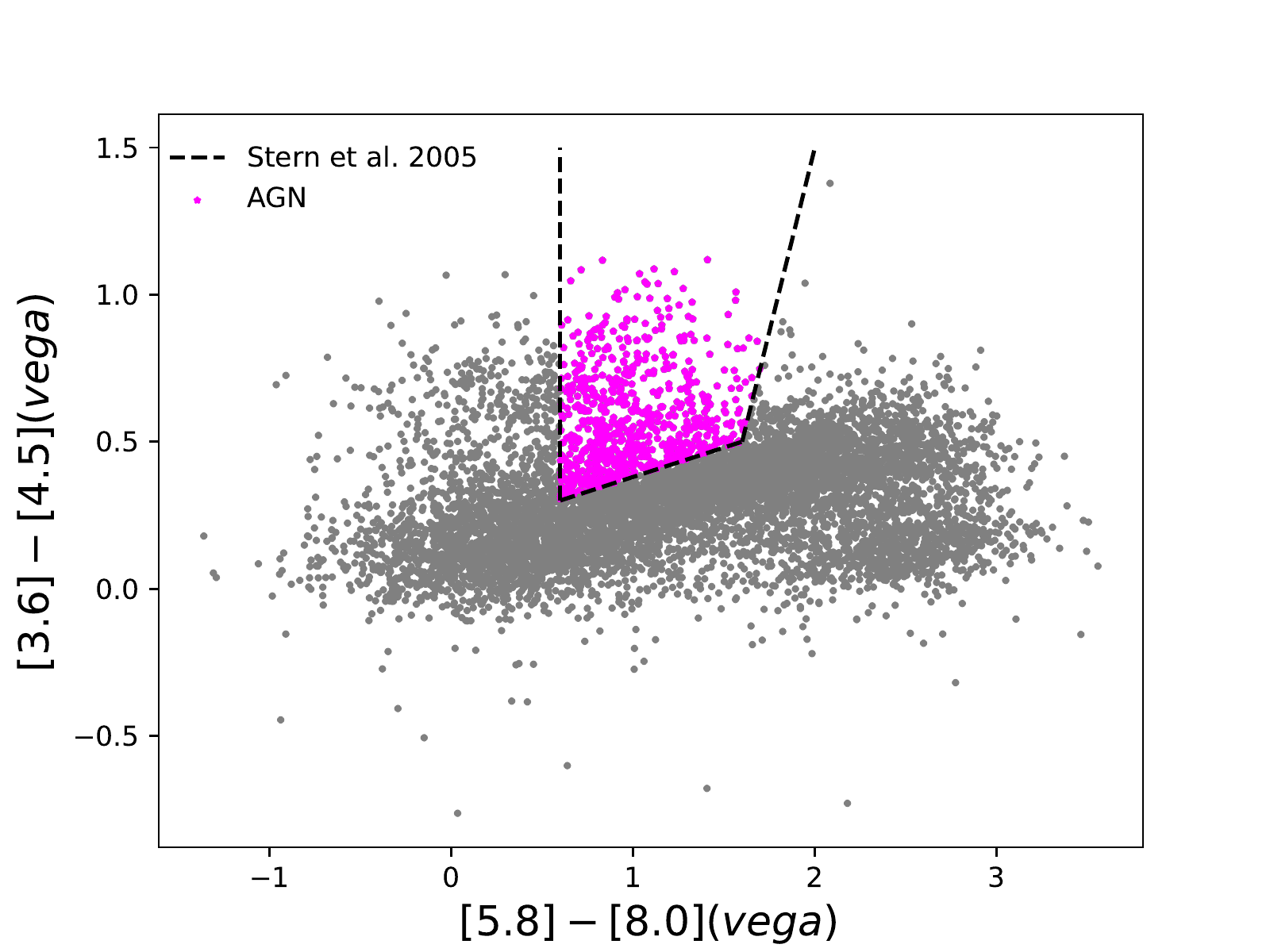}%
    \includegraphics[width=0.4\hsize]{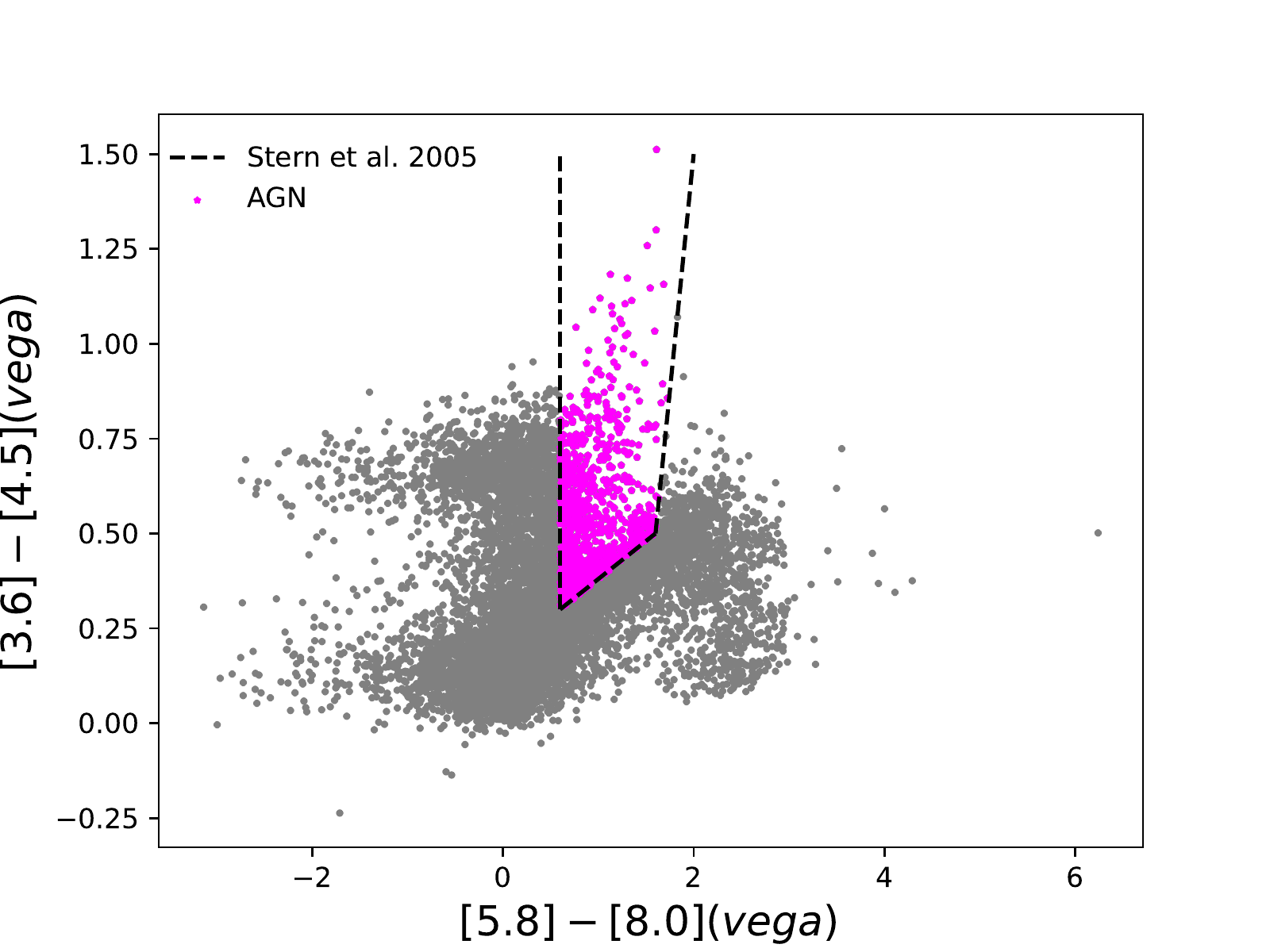}
    \caption{IRAC colour-colour diagrams for \EN (left column) and COSMOS (right column) fields. The AGN (magenta points) are selected following the criteria described in \citealt{Donley2012} (upper row black line) and \citealt{Stern2005} (lower row black line).}
    \label{fig:appendixAGN}
\end{figure*}

\begin{figure}[h]
    \centering
    \includegraphics[width=1\hsize]{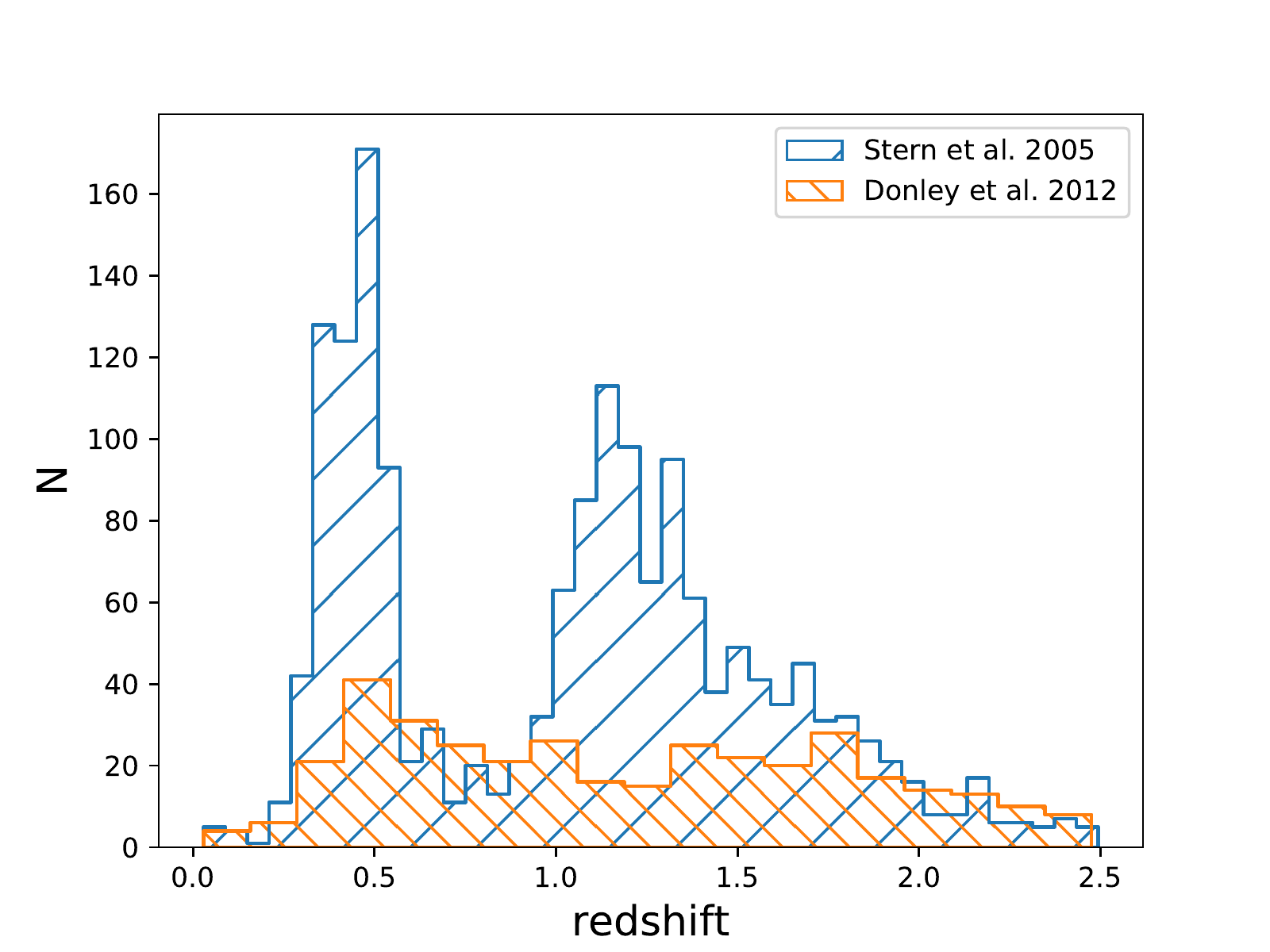}
    \caption{AGN redshifts distribution for the full sample of \EN+COSMOS fields.}
    \label{fig:appendixz}
\end{figure}

The AGN selection in \EN and COSMOS fields. 
We used two different selection criteria, based on the MIR features (IRAC bands) analysis:  \cite{Stern2005} and \cite{Donley2012}. 
Figure~\ref{fig:appendixAGN} shows the IRAC colour-colour selection using \cite{Donley2012} (upper panels) and \cite{Stern2005} (lower panels) methods. The redshift distribution of selected AGNs is shown in Fig.~\ref{fig:appendixz}.

\section{GALEX data simulation}

\begin{figure}[H]
    \centering
    \includegraphics[width=1\hsize]{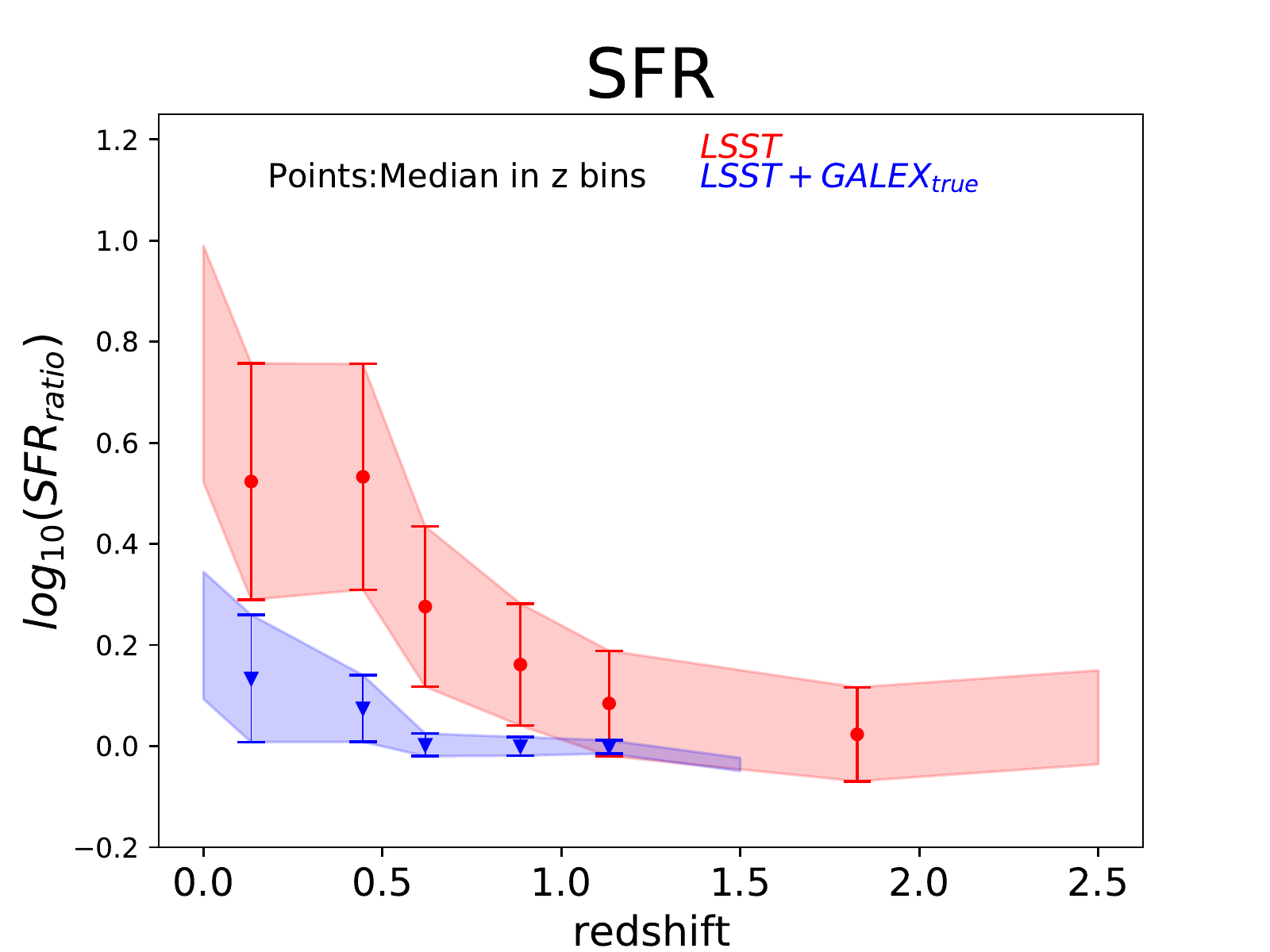}
    \caption{SFR ratio, defined as in Fig. \ref{fig:Density}, estimated with LSST data only (red) and LSST plus simulated GALEX observations ($GALEX_{true}$, in blue). Differently from the result shown in \ref{fig:SFRratio_IRAC}, in this case we are able to correct the overestimation, underlining the impact that high quality UV observations would have on the SFR estimate, when available for the entire sample.}
    \label{fig:galex_corr}
\end{figure}

\section{Scatter of the MS} 

\label{app:scatter}
 
\begin{figure}[h]
    \centering
    \includegraphics[width=1\hsize]{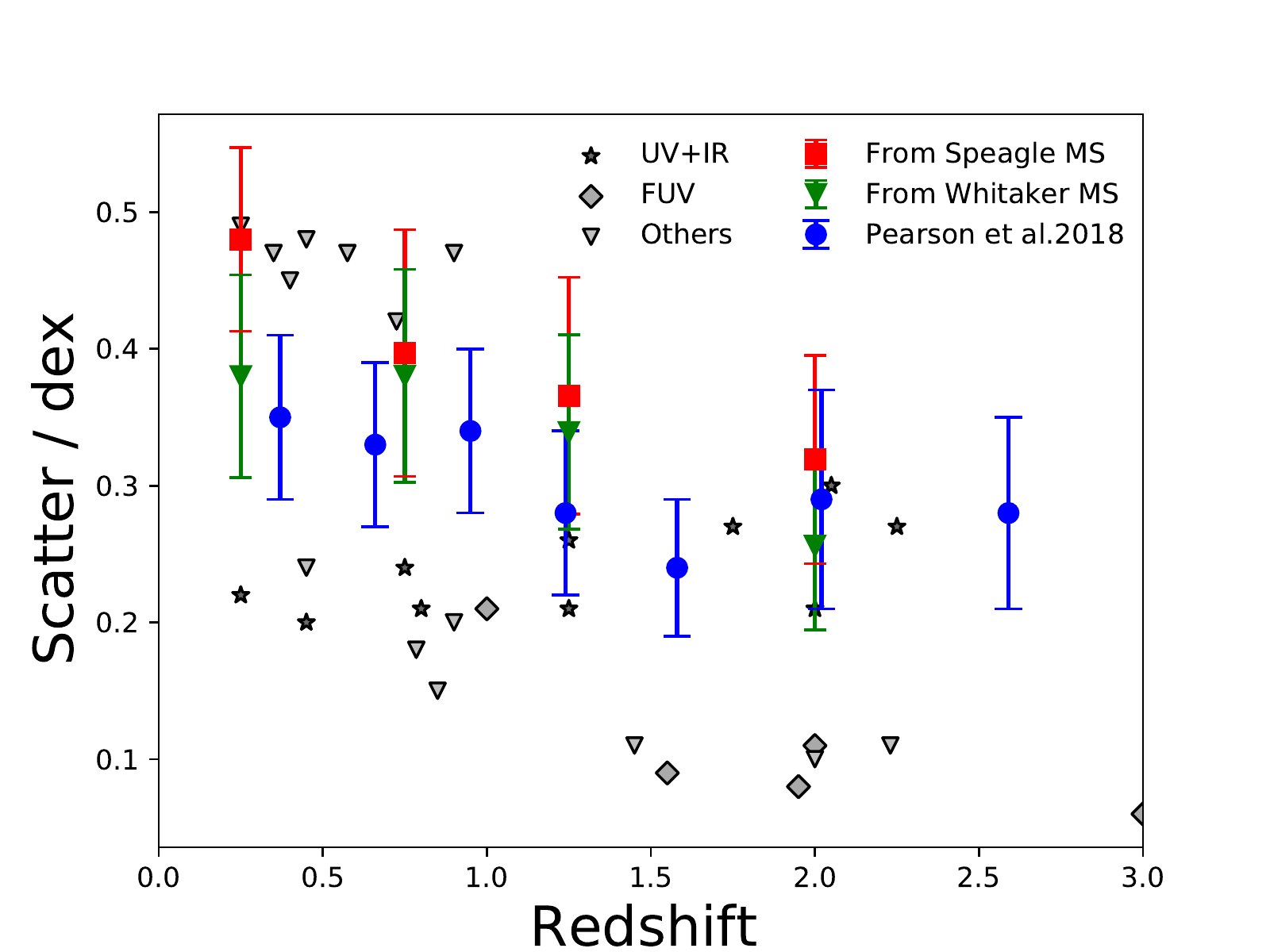}
    \caption{ Intrinsic scatter of our objects around exemplar MS, in comparison with the results found in previous works. Red and green points represent the scatter of our objects from \cite{Speagle2014} and \cite{Whitaker2017} MS respectively, blue points represent the results found in \cite{Pearson2018}, while grey stars, triangles and diamonds represent the results presented in \cite{Speagle2014}.}
    \label{fig:scatter}
\end{figure}

Star-forming galaxies follow a relatively tight, almost linear relation between SFR and \mstar known as MS. One of the most noticeable feature is that the MS relation at any given redshift shows a rather small scatter of $\sigma_{MS}$ that can vary from $\sim$ 0.2 to $\sim$ 0.4 dex (\citealt{Whitaker2012}, \cite{Speagle2014}, \citealt{Pearson2018}). Here we discuss the scatter of our sample from the MS, given that an high scatter could lead to an incorrect estimate of the physical parameters. Figure~\ref{fig:scatter} shows the scatter of our objects from two reference main sequence laws (\citealt{Speagle2014}, \citealt{Whitaker2017}) and we compare the results with the MS intrinsic scatter in the literature (\citealt{Speagle2014}, \citealt{Pearson2018}). We can notice that the scatter found in our sample is in agreement with those found previously in 
literature, within the error bars. This is valid also for the lowest redshift bin where the scatter appears to be the largest. Therefore, we are confident that we are  working with MS objects and that the input parameters are adequate to provide us with reliable physical properties for the purposes of this analysis. The origin of this scatter can be traced back to different enhancements or decrements events of star formation that could occur during the galaxy life-time.
In fact, large scale gas inflow/outflow events can trigger gas compaction/depletion phenomena, that can lead  to an enhancement/decrement of the SFR of the galaxy (\citealt{Tacchella2016}).

\end{appendix}

%
%

\end{document}